\documentclass{egpubl}

\JournalSubmission    
 \electronicVersion 

\ifpdf \usepackage[pdftex]{graphicx} \pdfcompresslevel=9

\else \usepackage[dvips]{graphicx} \fi

\PrintedOrElectronic

\usepackage{t1enc,dfadobe}

\usepackage{egweblnk}
\usepackage{cite}
\usepackage{multirow}
\usepackage{subfig}
\usepackage{caption}

\title[Image-based remapping of spatially-varying material appearance]%
      {Image-based remapping of spatially-varying material appearance}

\author[A. Sztrajman, J. K\v{r}iv\'{a}nek, A. Wilkie, T. Weyrich]
{Alejandro Sztrajman$^1$, Jaroslav K\v{r}iv\'{a}nek$^2$, Alexander Wilkie$^2$, Tim Weyrich$^1$
 \\
 $^1$University College London, United Kingdom.\\
 $^2$Charles University, Czech Republic.
 }

\usepackage{calc}
\usepackage{printlen}
\usepackage{lipsum}

\usepackage{microtype}
\usepackage{amsmath}

\usepackage[usenames,dvipsnames]{xcolor}
\usepackage[normalem]{ulem}  
\def\clap#1{\hbox to 0pt{\hss #1\hss}}%
\def\initials#1{\protect\clap{\smash{\raisebox{1.4ex}{\tiny{\textsf{\textit{#1}}}}}}}%
\makeatletter
\newcommand{\NOTE}[3]{\protect\@ifundefined{hidecomments}{%
  \strut{\color{#2}{\hspace{0pt}\initials{#1}\protect{[#3]}}}%
  }{}}
\newcommand{\EDITbyauthor}[4][]{\protect\@ifundefined{hidecomments}{%
  \strut{\color{#3}{\hspace{0pt}\initials{#2}\protect\sout{#1}{ #4}}}%
  }{}}
\newcommand{\EDITredandgreen}[4][]{\protect\@ifundefined{hidecomments}{%
  \strut{\color{Red}{\hspace{0pt}\protect\sout{#1}{\color{ForestGreen}{#4}}}}%
  }{}}
\newcommand{\EDITgreenonly}[4][]{\protect\@ifundefined{hidecomments}{%
  \strut{\color{ForestGreen}{#4}}%
  }{}}

\newcommand{\NOTEboxed}[3]{\protect\@ifundefined{hidecomments}{%
  {\centering\fbox{\parbox{0.97\linewidth}{\protect\EDIT{#1}{#2}{#3}}}}%
  }{}}
\newcommand{\EDITPRESERVEbyauthor}[4][]{\protect\@ifundefined{hidechanges}{%
  \strut{\color{#3}{\hspace{0pt}\initials{#2}\protect\sout{#1}{ #4}}}%
  }{\strut{ #4}}}

\newcommand{\EDIT}[4][]{\EDITbyauthor[#1]{#2}{#3}{#4}}
\newcommand{\EDITPRESERVE}[4][]{\EDITPRESERVEbyauthor[#1]{#2}{#3}{#4}}

\makeatother
\newcommand{\TWedit}[2][]{\protect\EDIT[#1]{TW}{Orange}{#2}}

\newcommand{\AS}[1]{\protect\NOTE{AS}{Red}{#1}}

\newcommand{\ASadd}[2][]{\protect\EDITPRESERVE[#1]{AS}{Red!45!Black}{#2}}
\def\ignore#1{}

\usepackage{bm}

\begin{document}

\teaser{
  \centering
  \includegraphics[height=4.5cm]{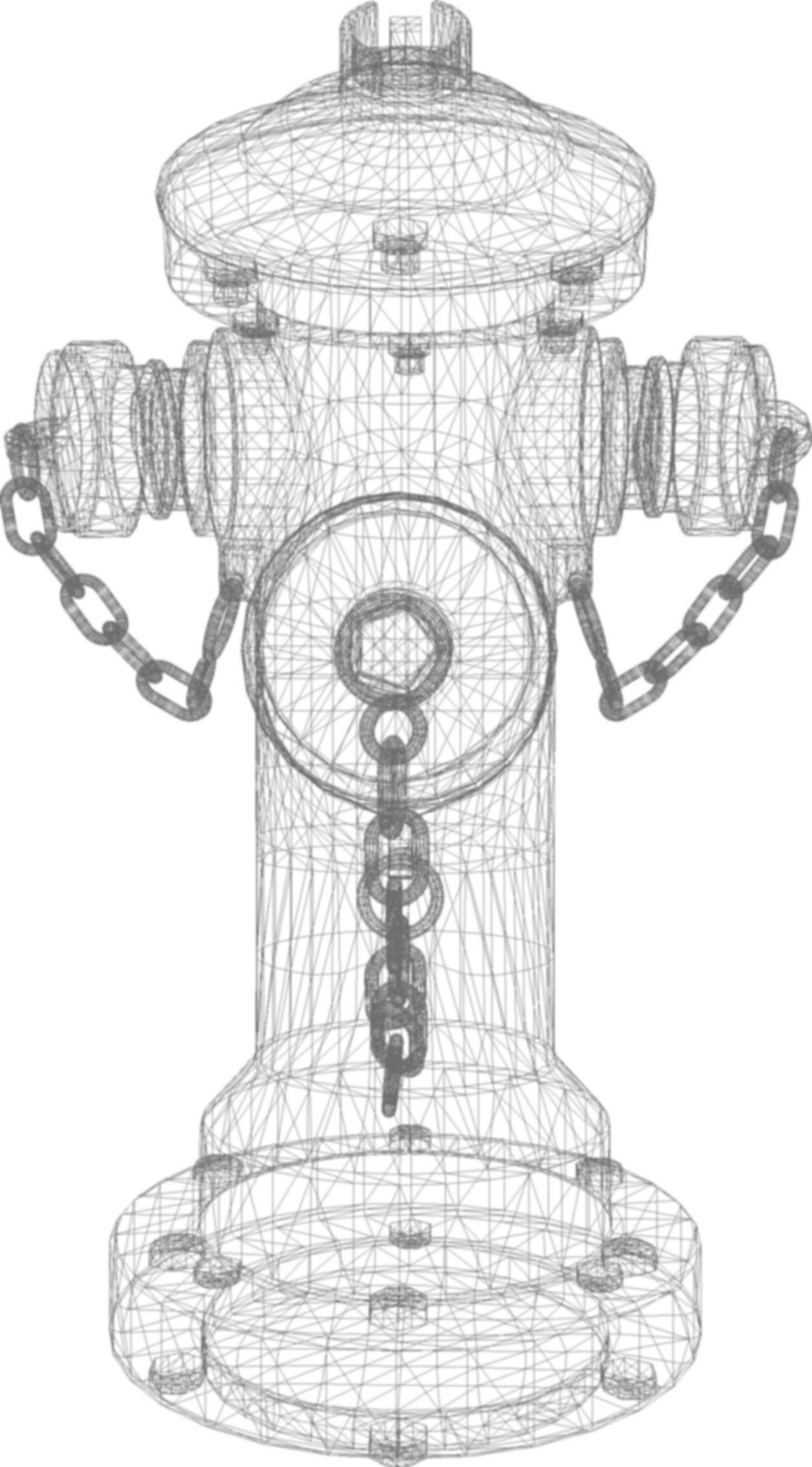}
  \includegraphics[height=4.5cm]{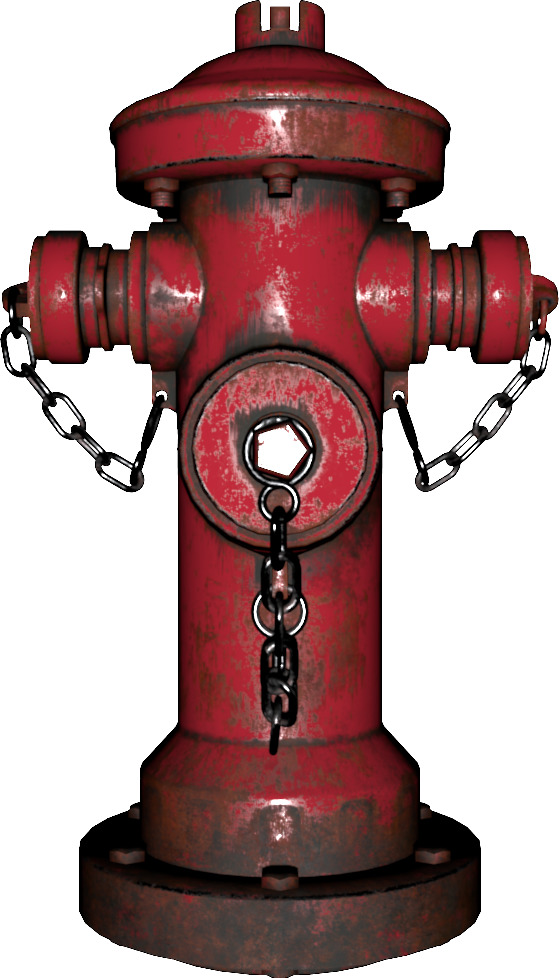}
  \includegraphics[height=4.5cm]{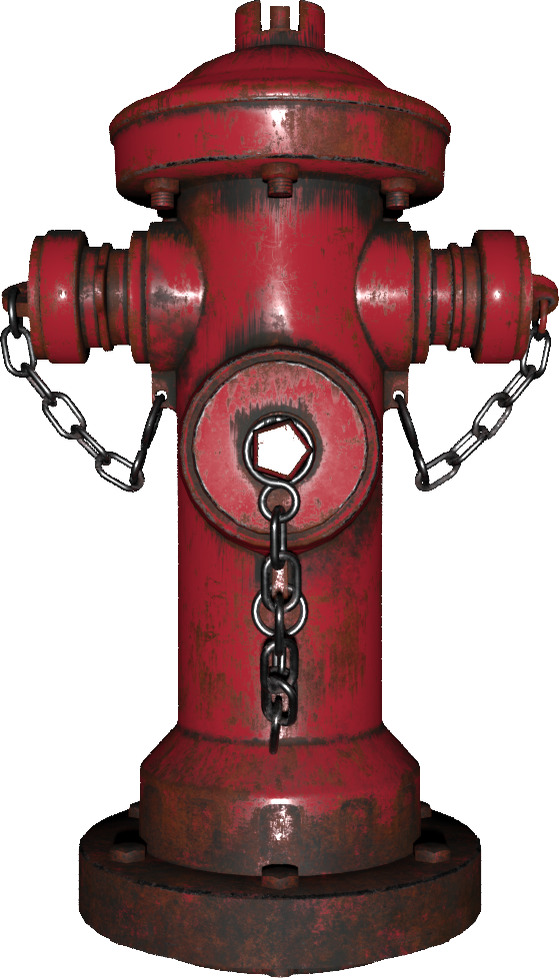}
  \caption{\label{fig:teaser}%
    Remapping of material from Blender-Ward (center) to Cycles-GGX (right).\vspace{2ex}%
  }
}

\maketitle
\begin{abstract}
BRDF models are ubiquitous tools for the representation of material appearance. 
However, there is now an astonishingly large number of different models 
in practical use. Both a lack of BRDF model standardisation across 
implementations found in different renderers, as well as the often semantically 
different capabilities of various models, have grown to be a major hindrance to 
the interchange of production assets between different rendering systems. 
Current attempts to solve this problem rely on manually finding visual 
similarities between models, or mathematical ones between their functional 
shapes, which requires access to the shader implementation, usually unavailable 
in commercial renderers. We present a method for automatic translation of 
material appearance between different BRDF models, which uses an image-based 
metric for appearance comparison, and that delegates the interaction with the 
model to the renderer. We analyse the performance of the method, both with 
respect to robustness and visual differences of the fits for multiple 
combinations of BRDF models. While it is effective for individual BRDFs, the 
computational cost does not scale well for spatially-varying BRDFs.
Therefore, we further present a parametric regression scheme that approximates
the shape of the transformation function and generates a reduced representation
which evaluates instantly and without further interaction with the renderer.
We present respective visual comparisons of the remapped SVBRDF models for
commonly used renderers and shading models, and show that our approach is able to
extrapolate transformed BRDF parameters better than other complex regression schemes.
\end{abstract}

\section{Introduction}

Computer-generated imagery workflows commonly involve a broad range of
modelling and rendering tools, each targeting different goals and 
requirements~\cite{schregle2013}, and the exchange of data between these tools 
is hindered by incompatible representations. As a consequence, existing model 
assets frequently have to be redesigned to be used in other software, resulting 
in large modelling overheads. This is particularly true in the case of material
models. So far, a great number of BRDF models has been developed for appearance 
representation, but a lack of a standardisation and renderer-specific implementation 
details lead to visual deviations even between identically named reflectance models.

The current abundance of BRDF models reflects that no single model is
able to realistically reproduce the full range of available measured
materials~\cite{brady2014genbrdf,guarnera2016brdf}. However, for a
given material, represented using one model, it is often possible to
find a new set of parameters which approximates its appearance with a
different model. Many existing rendering systems support such
\textit{remapping} of material parameters to address the incompatibility
between models, and be it to remain backwards-compatible to older version
of their software~\cite{pharr2016pbr,corona2017}. That said, these remappings
are often based on manually determined, or heuristic relations between
the functional shapes of the models,
\TWedit[which requires access to the model implementations,]{}%
or incur oversimplifications by assuming one-to-one correspondence
between individual parameters of both models. The problem is
exacerbated by popular renderers and graphics engines using their
own shading models~\cite{maya2017,unreal2017,unity2017}.

An automatic solution to this problem needs to consider the constraints of the
real-world scenario where a material is interchanged between different third-party renderers.
In this situation, we do not have access to the implementation of the shaders,
only to the model parameters and the resulting renderings. To address this problem,
we present an image-based method for the remapping of BRDFs which works for
closed-source renderers, assuming no knowledge of the model implementations.
We analyse the robustness of the method applied to a set of BRDF models, and we
discuss common issues and strategies to improve the stability of the method
in different types of materials.

In addition, we present a \ASadd{regression scheme} to generate a reduced representation
of the transformation which evaluates instantly and without further interaction with the
renderer, allowing the fast remapping of entire parameter texture maps. We show visual
comparisons of the remapping of spatially-varying BRDF models (SVBRDF) that illustrate
the ability of our approach to provide a close match between different renderers, even
when remapping between very different shading models.

\section{Related Work}

\subsection{Reflectance Remapping}

Traditionally, appearance modelling dealt with finding reflectance
models that would agree well with measured data. Accordingly, a large
body of work on fitting of reflectance models
exists~\cite{marschner1998inverse,weyrich2008principles,guarnera2016brdf}.
In contrast, little academic attention has been paid to the direct
translation between BRDF models.
Within commercial products, the arguably most prominent software to
remap reflectance from one model to another is Allegorithmic's
Substance Painter~\cite{allegorithmic2017},
a dedicated tool to author appearance for a wide range of target
platforms. In order to address the variability in renderer-specific
BRDF models and implementations, they contain various export functions
that employ manually optimised heuristics to remap BRDF parameters for
specific target rendering engines. Manual creation of such heuristics,
however, can be costly and does not necessarily lead to optimal
results~\cite{cyrille.comm}.

Another example would be renderers that remap reflectance from older
versions' legacy representations, such as PBRT~\cite{pharr2016pbr} or
Corona~\cite{corona2017}. The latter switched from a variant of the
Ashikhmin-Shirley BRDF~\cite{ashikhmin2000anisotropic} to a GGX microfacet
BRDF~\cite{walter2007microfacet,burley2012physically} and remaps BRDF
specifications by analytically matching the width of the models' specular
lobes~\cite{jaroslav.comm}.

\subsection{Appearance Comparison}
Quantifying (dis)similarity between two BRDFs is a problem encountered in any BRDF
fitting work, and our remapping is no exception. Ngan et al.~\cite{ngan2005} follow
Lafortune et al.~\cite{lafortune1997} and employ a simple $L_2$ distance between
cosine-weighted BRDF values. We find that such a metric puts disproportionate emphasis
on matching BRDF peaks at the expense of tails, which can result in appearance deviations.

In their follow-up work, Ngan et al.~\cite{ngan2006image} argue for an image-based
metric, where the dissimilarity between two BRDFs is modelled as the difference
between the rendered images with the respective BRDFs under natural illumination.
Recently, Havran et al.~\cite{havran2016} confirmed the validity of the image-based
methodology through psychophysical experiments and, furthermore, designed
specialised geometries that provide richer information on material properties
than the simple sphere used by Ngan et al.
We follow this image-based strategy for two reasons: first, it has been repeatedly
shown to correlate well with the perceived material differences; second, our setup
lends itself well to rendering images using any (unknown) BRDF, whereas obtaining
individual BRDF values using an off-the-shelf renderer may be more difficult.

We focus our efforts on finding a remapping scheme which results in parameters
that vary smoothly with respect to changes in the source material parameters.
For uniform BRDFs this is an expected behaviour of the transformation, and
deviations are suggestive of problems in the optimisation, such as finding
local minima or output parameters that only look similar in a particular scene setting.
For spatially-varying (SV)BRDFs, the smoothness of the mapping is even more
important because the material parameters across the surface are computed
by interpolation. If the parameters vary abruptly this is likely to produce
wrong SVBRDF remappings, even if the appearance of each individual texel is
correctly matched.

\section{Remapping of Uniform Materials}
\begin{figure}[htb]
  \centering
  \includegraphics[width=.85\linewidth]{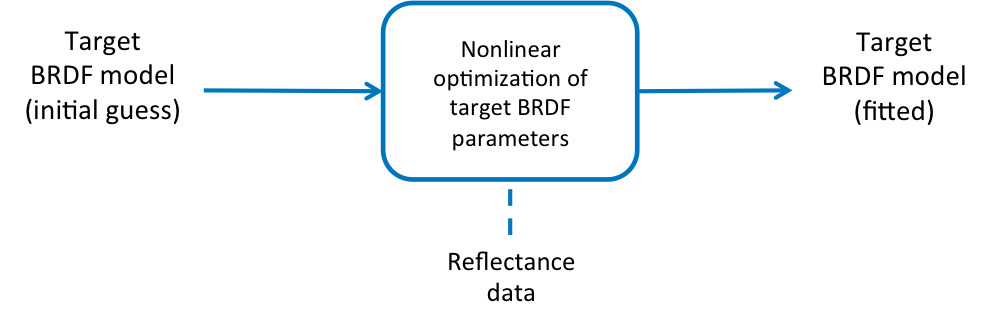}
  \caption{\label{fig:scheme-fitting}%
    Broad scheme for BRDF fitting.}
\end{figure}

The process of BRDF remapping is similar in structure to the fitting of BRDFs,
where we start with a \emph{target} BRDF model and an initial guess of the
parameters, and we want to fit reflectance data, generally measured from
a real-world material. This involves the minimisation of the difference
between the appearances of the BRDF and the data, performed through
non-linear optimisation of the parameters of the BRDF model
(Figure~\ref{fig:scheme-fitting}).\\

\begin{figure}[htb]
  \centering
  \includegraphics[width=.85\linewidth]{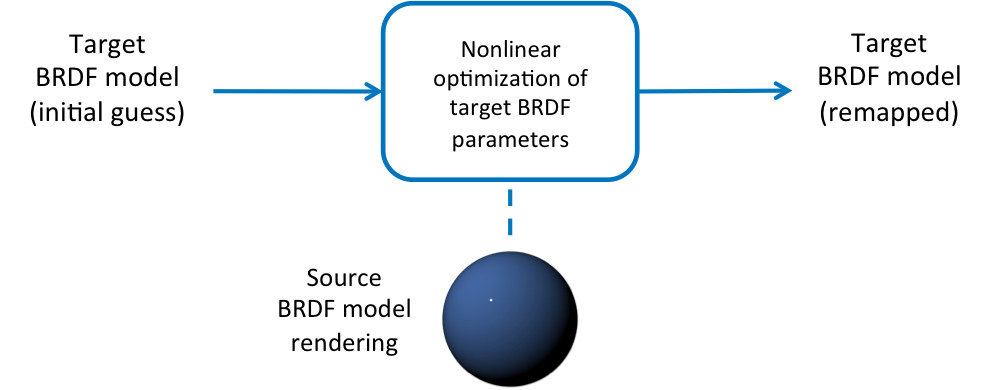}
  \caption{\label{fig:scheme-remapping}%
    BRDF remapping scheme.}
\end{figure}

In the case of BRDF remapping (Figure~\ref{fig:scheme-remapping}) the scheme is
analogous; however, instead of using measured data we are now matching the appearance
of the target BRDF model with another \textit{source} BRDF model. In this scheme we
assume no direct access to the implementation of the BRDF models.
In particular, the target model is assumed to belong to an external renderer
in a typical usage scenario of our technique. In order to perform an appearance
comparison under these conditions, we measure the difference in image space, by
comparing rendered images of a single scene using each of the two BRDF models.\\

Thus, in each step of the optimisation we only need to be able to generate new renders
of this scene with the target BRDF. The image difference is then computed with an $L_2$
metric in colour space, which is common practice in the context of BRDF
fitting~\cite{ngan2006image} (other metrics are used as well, but no single distance
metric has emerged as superior choice for general BRDF fitting.)

In the remainder, we will consider three optimisation strategies to remap a (uniform)
BRDF specification to parameters of a different model. Section~\ref{sec:svbrdf} will
then present our approach to extend the remapping to spatially-varying appearance.

\subsection{Optimisation Strategies}
\label{sec:uniform-strategies}

A simple optimisation scheme which attempts to fit all model parameters at once
(as shown in Figure~\ref{fig:scheme-remapping}) often leads to local minima during
the optimisation, due to the coupling between the diffuse and specular terms in the model.
In Section~\ref{sec:results:uniform} we provide a systematic analysis of the stability
of the remapping scheme. In order to improve the stability of the optimisation we test
the following two variants of our remapping scheme.

\subsubsection*{Two-stage remapping}
\begin{figure}[htb]
  \centering
  \includegraphics[width=.95\linewidth]{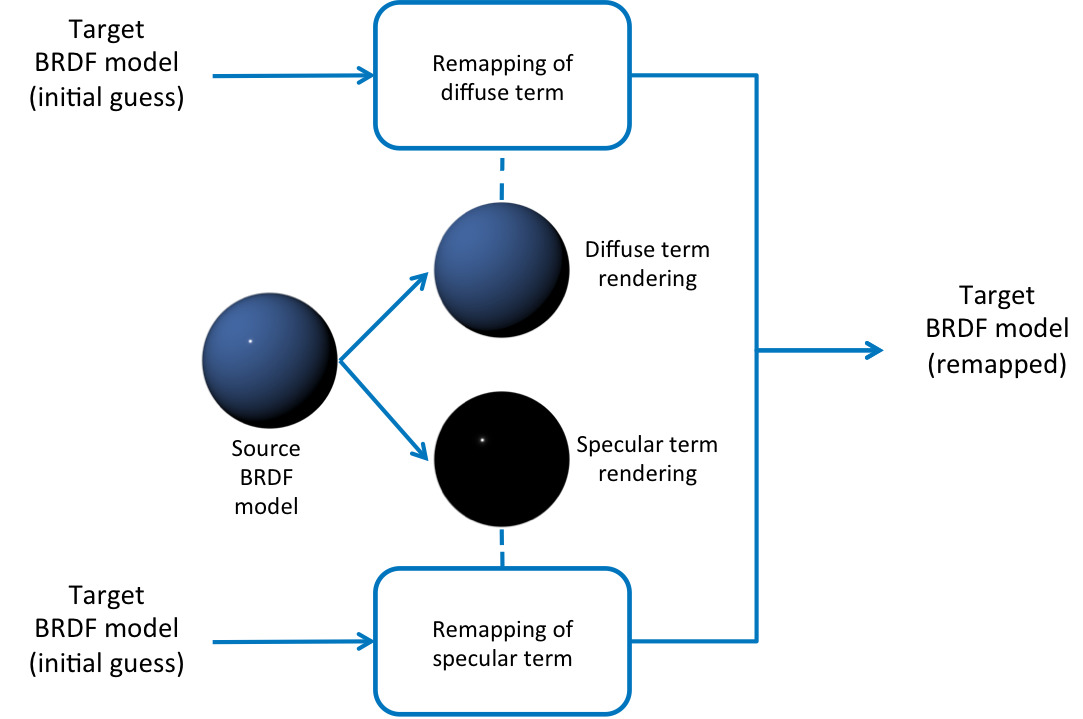}
  \caption{\label{fig:scheme-remapping-2stage}%
    BRDF remapping scheme in two stages. Diffuse and specular components are remapped independently.}
\end{figure}

In this scheme the diffuse and specular terms are remapped independently
(Figure~\ref{fig:scheme-remapping-2stage}).
This is not unlike BRDF fitting to real-word data where diffuse and
specular reflectance may be separated optically~\cite{debevec2000acquiring}
or statistically~\cite{weyrich2006analysis} before conducting separate fits.
In our case, it requires source renderings of the diffuse-only, and purely
specular components, respectively. In the end we obtain remapped versions
of each term which are merged in the remapped target BRDF model.\\

\subsubsection*{Three-stage remapping}
\begin{figure}[htb]
  \centering
  \includegraphics[width=.95\linewidth]{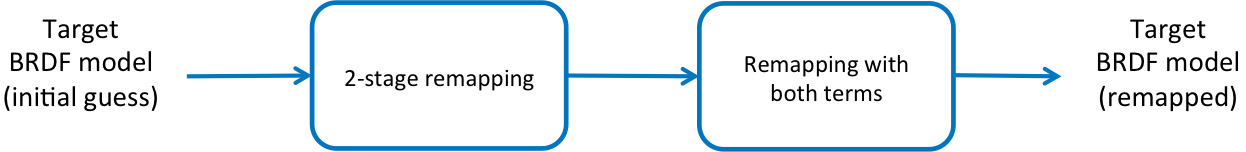}
  \caption{\label{fig:scheme-remapping-3stage}%
    BRDF remapping scheme in three stages.}
\end{figure}

The two-stage remapping assumes an independence of diffuse and specular terms that might
not hold true for some layered materials. The three-stage scheme
(Figure~\ref{fig:scheme-remapping-3stage}) recovers the coupling between both terms by
using the results of the two-stage scheme as a good starting guess for a subsequent
simple remapping that optimises all parameters simultaneously, reducing the chance of falling
into local minima.\\

\section{Analysis of Uniform Material Remapping}
\label{sec:results}

We tested our approach using three renderers: Mitsuba; Blender's internal 
preview renderer; and Cycles, a physically-based renderer that is currently the 
most commonly used off-line renderer in Blender. Due to differences in how light
source intensities are specified across renderers, we further had to match irradiance 
before remapping an external BRDF to a Mitsuba BRDF. This was done by a global 
scale determined from the ratio of diffuse-only renderings from the two  
renderers.

Our uniform BRDF remapping code uses Mitsuba in its inner loop, and thus the 
optimisation can take source BRDFs from arbitrary renderers while the target 
BRDF has to be from within Mitsuba. In Section~\ref{sec:results-svbrdf-remapping} we 
will show that this does not represent a limitation for the remapping, since the 
transformation to Mitsuba can be used as an intermediate step in a sequence of 
remappings.

\subsection{Uniform Fitting Strategies}
\label{sec:results:uniform}

We begin by evaluating the three optimisation strategies for remapping of
uniform BRDFs which were introduced in Section~\ref{sec:uniform-strategies}. We 
performed a systematic study of these remapping schemes via an analysis of the 
robustness of the transformation that links the parameters of the models. We did 
this for multiple combinations of BRDF models that are available in 
Mitsuba~\cite{mitsuba} (Ashikhmin-Shirley, Beckmann, GGX, Phong, Ward). 
For the sake of brevity, we focus here only on a few of these combinations, to  
demonstrate a few common effects we encountered when dealing with  
remapping between different BRDFs within Mitsuba.

We used a simple scene for the image-based appearance comparison: a sphere of radius
2 located in the origin illuminated by a point lightsource. This produces a sampling of
only a two-dimensional slice of the BRDF space, that depends on the relative position of
the illumination source. Although we find that the resulting transformations of parameter
are not highly dependent on the choice of light position, in
Section~\ref{sec:results:nonheadlight} we will discuss its effect on the visual match between
BRDF models.

The renderings were generated as colour space HDR images ($512\times512$),
and the non-linear optimisation was performed using the Trusted Region Reflective
method, enforcing positive values on the remapped parameters.

\subsubsection{Conductors}

\begin{figure}[htb]
  \centering
  \includegraphics[width=.95\linewidth]{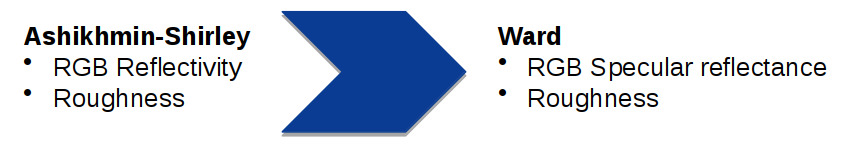}
  \includegraphics[width=.95\linewidth]{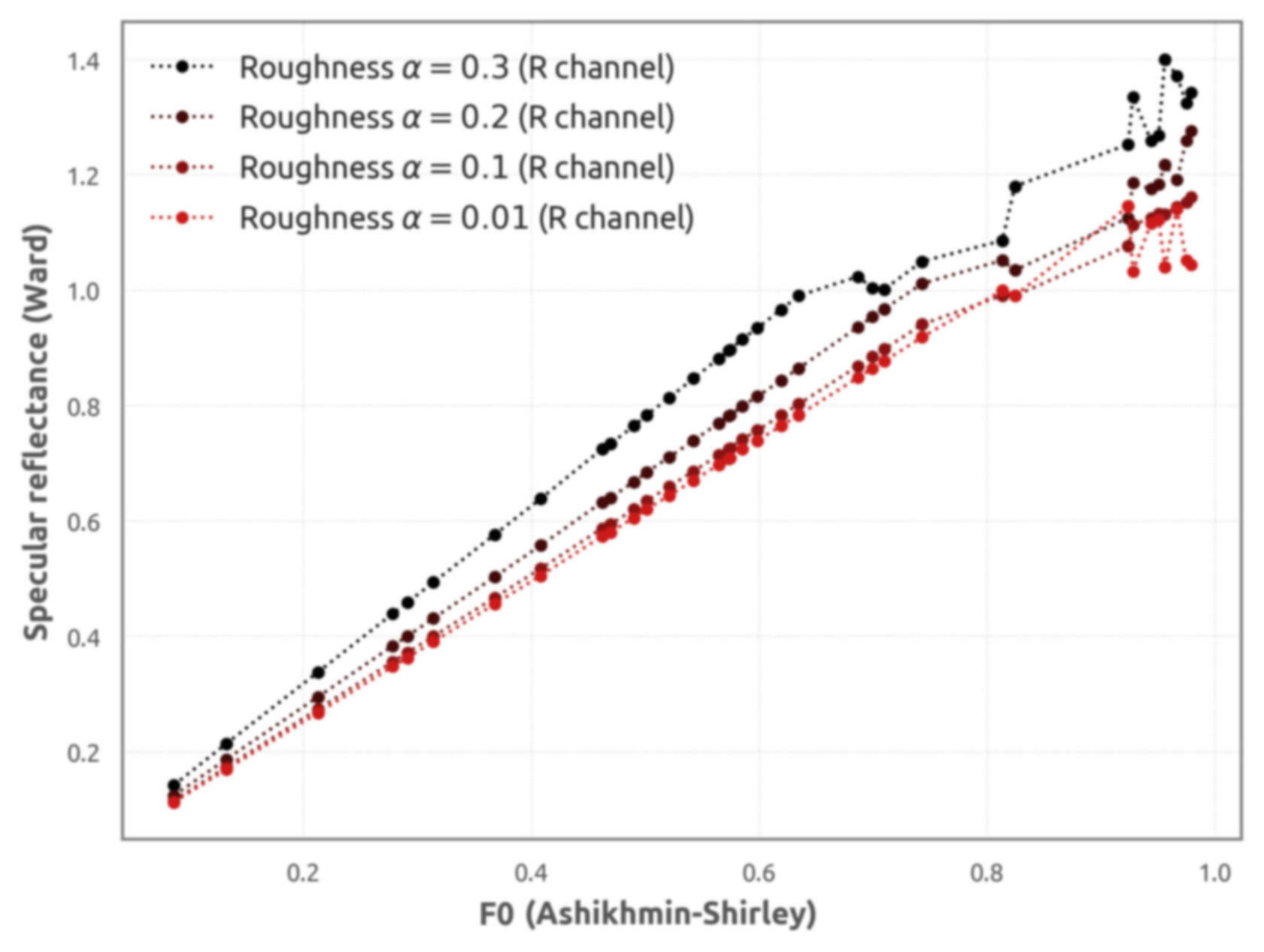}
  \caption{\label{fig:as2ward-conductors-channelR-finalplot}%
    Remapping of conductors from Ashikhmin-Shirley to Ward. Detail of parameters 
    in Mitsuba, and plot of specular reflectance (Ward) vs Fresnel coefficient (AS).}
\end{figure}

Our analysis of the remapping of conductors comprises 60 materials from 
Mitsuba's database. In Figure~\ref{fig:as2ward-conductors-channelR-finalplot} we 
show the results for a remapping from Ashikhmin-Shirley (source) to Ward 
(target). In these implementations the specular terms in both models are 
described by an RGB specular parameter and the roughness (single-channel), 
essentially characterising the intensity and the spread of the lobes. We show 
the remapping of a specular parameter in Ashikhmin-Shirley to an analogous 
parameter in Ward (both single-channel), for multiple fixed values of roughness.
The information provided by the index of refraction in the standard BRDF 
interface in Mitsuba is here condensed into the Fresnel coefficient $F_0$, which can 
be expressed in its more general form as:
\begin{equation}
F_0 = \frac{(c-1)(c^*-1)}{(c+1)(c^*+1)}
\label{eq:f0}
\end{equation}
where $c$ is the complex index of refraction.

The expected output of the remapping is a smoothly varying correspondence 
between source and target parameters. In a typical usage case, the user would 
expect small changes in the source material to correspond to small changes in 
the exported appearance. We will show that a deviation from this behaviour 
usually signals a decreasing capacity of the target model to match the source, 
or the occurrence of local minima during the optimisation. The stability of 
the transformation will prove crucial when we deal with the remapping of 
spatially-varying BRDFs (Section~\ref{sec:svbrdf}) which are 
reconstructed by interpolation of multiple uniform materials.

In Figure~\ref{fig:as2ward-conductors-channelR-finalplot} the transformation 
shows a smooth behaviour for most materials, but exhibits instabilities for 
parameters which are remapped to specular reflectance $>$ 1. These can be traced 
back to this particular implementation of Ward, where the values of specular 
reflectance are trimmed to avoid energy loss, and is representative of the 
implementation-dependent behaviour that we may find in renderers. Most of the 
instabilities we found during our study shared this behaviour of exhibiting 
well-localised regions in parameter space where the remapping becomes 
unreliable.

\begin{figure}[htb]
\centering
\includegraphics[width=.95\linewidth]{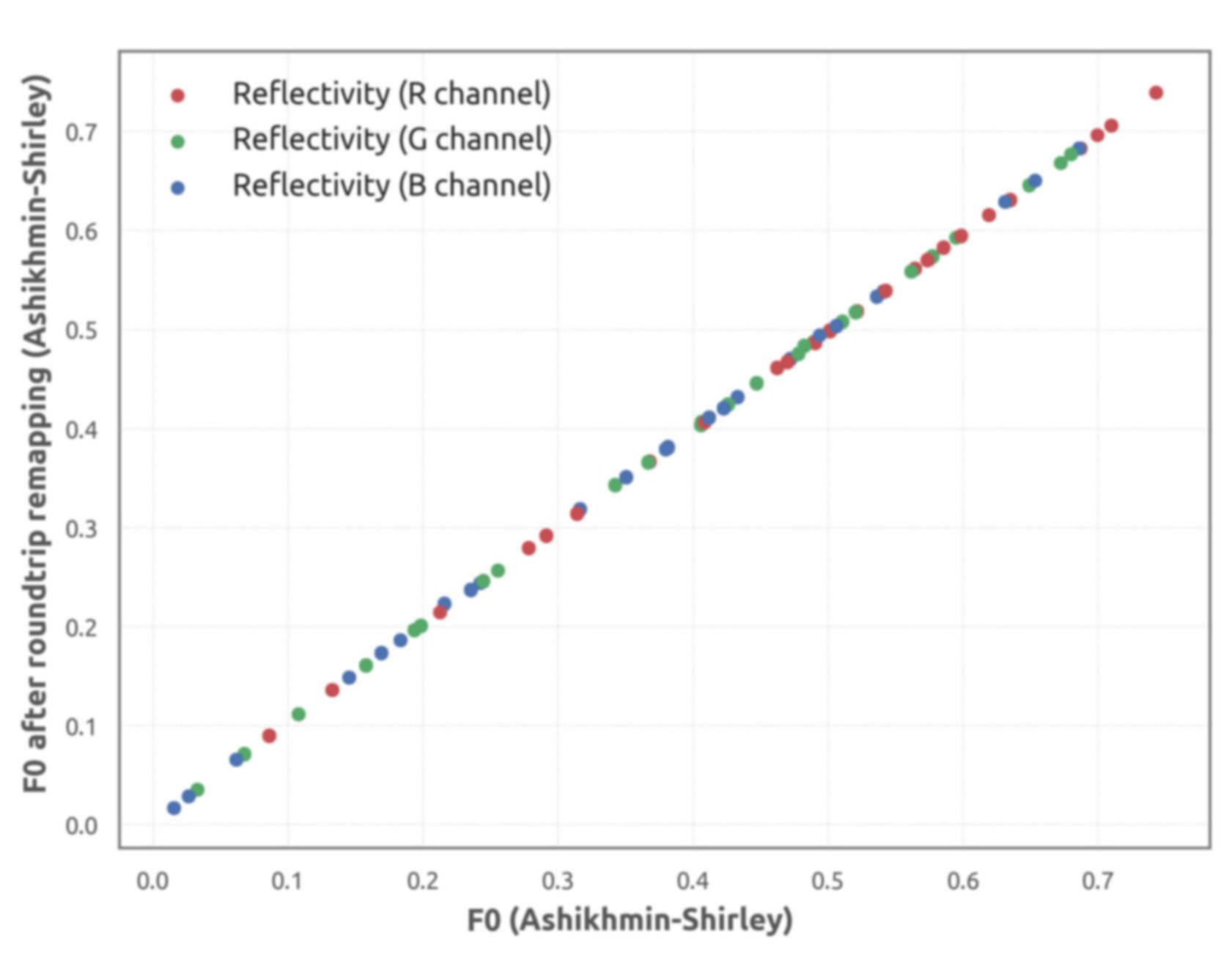}
\caption{\label{fig:as2ward2as-conductors-RGB-roundtrip-finalplot}%
  Round-trip remapping of conductors from Ashikhmin-Shirley to Ward and then 
  back to Ashikhmin-Shirley. Remapped Fresnel coefficient $F_0$ vs original $F_0$.}
\end{figure}

After filtering the unstable cases from 
Figure~\ref{fig:as2ward-conductors-channelR-finalplot}, 
Figure~\ref{fig:as2ward2as-conductors-RGB-roundtrip-finalplot} shows the
results for a round-trip remapping, where we transformed the parameters back 
to the initial Ashikhmin-Shirley model. The result is a straight line of unitary 
slope, which shows that in this particular case, the parameters go back to their 
original values after the two remappings. This speaks for the general robustness 
of the approach, and indicates that we generally can recover the 
original appearance after a remapping takes place.

\subsubsection{Dielectrics}
\label{sec:dielectrics}

\begin{figure}[hbt!]
  \centering
  \includegraphics[width=0.9\linewidth]{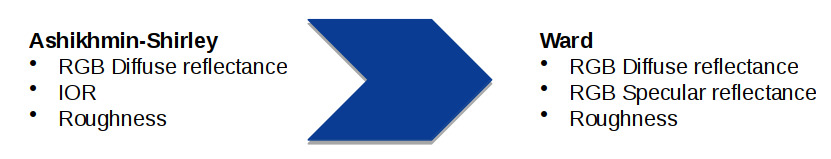}
  \includegraphics[width=.95\linewidth]{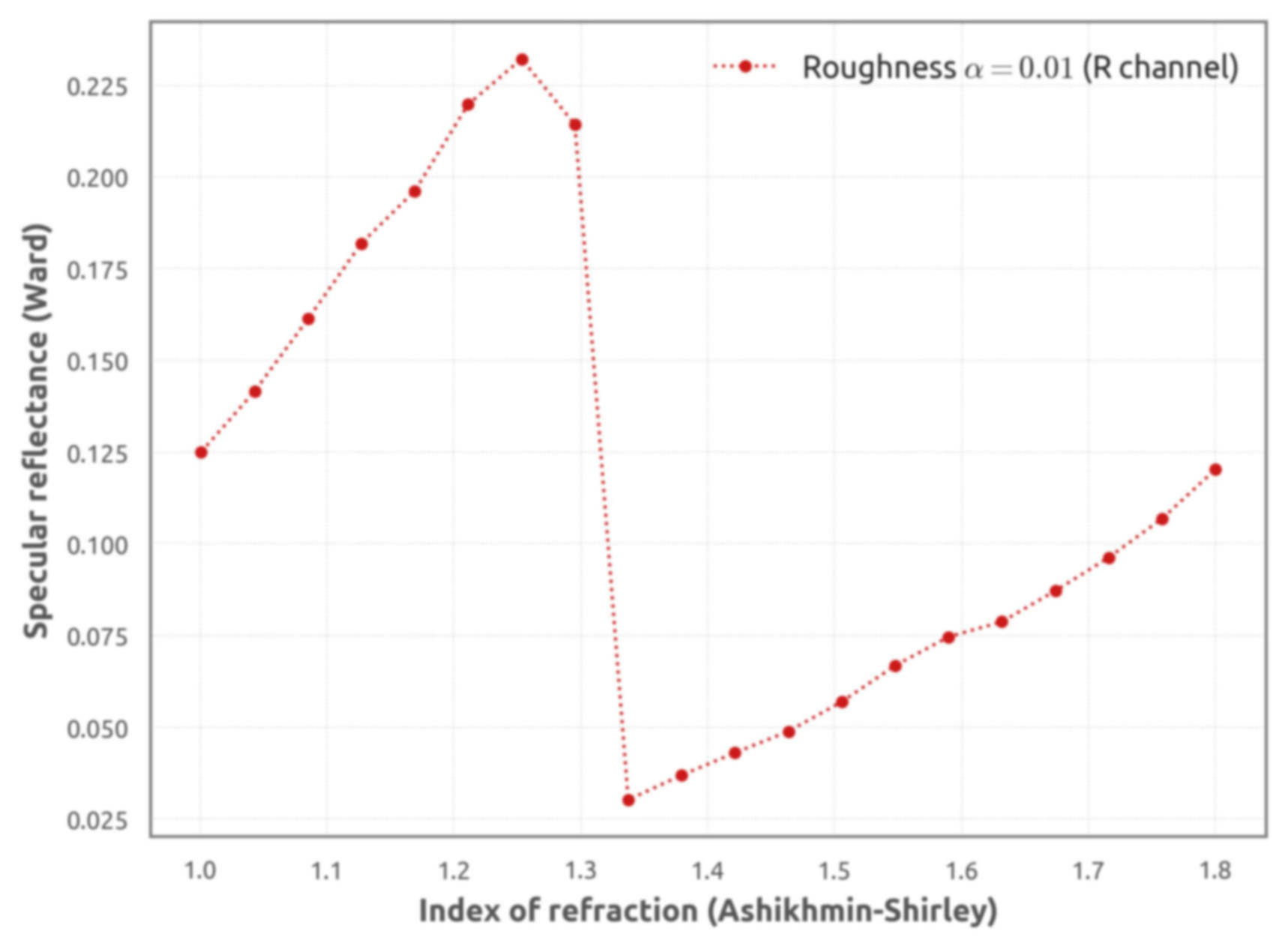}
  \caption{\label{fig:as2ward-plastics1stage-diffmat3-finalplot-onlyalpha001}%
    Simple remapping of dielectrics from Ashikhmin-Shirley to Ward. Detail of 
    parameters in Mitsuba for both models, and plot of specular reflectance 
    (Ward) vs IOR (AS).}
\end{figure}

The reflectance of dielectric materials includes an additional diffuse component.
The specular component has a similar behaviour for all channels, and is usually
approximated by a single parameter (e.g., the index of refraction).
In Figure~\ref{fig:as2ward-plastics1stage-diffmat3-finalplot-onlyalpha001}
we show the results of remapping from Ashikhmin-Shirley to Ward, using a simple 
optimisation with both diffuse and specular parameters. The plot corresponds to 
a parameter sweep of the IOR in Ashikhmin-Shirley for fixed diffuse and
roughness parameters.

\begin{figure}[ht]
\begin{center}
\begin{tabular}{cccp{2.5em}}
	\includegraphics[width=0.28\linewidth]{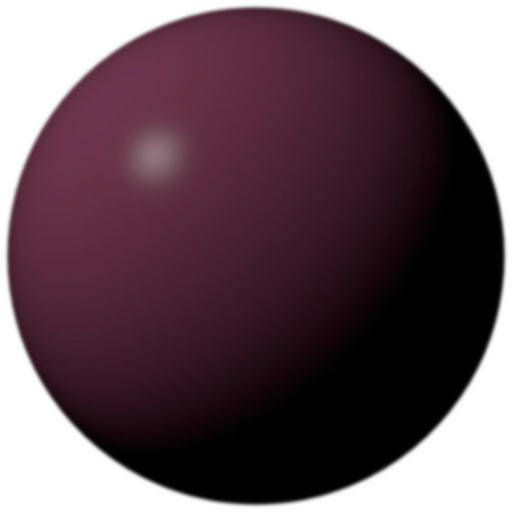} & \includegraphics[width=0.28\linewidth]{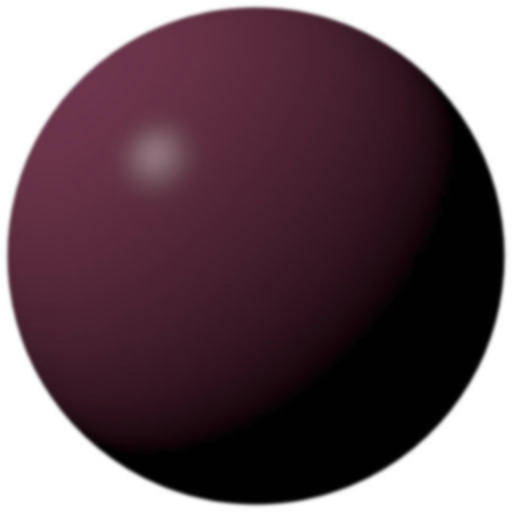} &
	\includegraphics[width=0.28\linewidth]{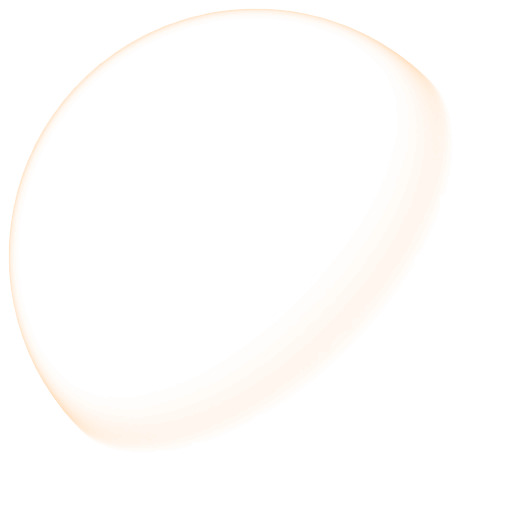} & \hspace*{1em}\rlap{\raisebox{1.5cm}{\multirow{2}{*}{\includegraphics[width=0.9\linewidth]{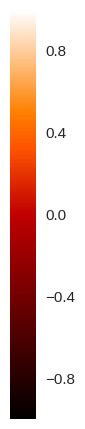}}}}\hspace*{2em} \\
    \includegraphics[width=0.28\linewidth]{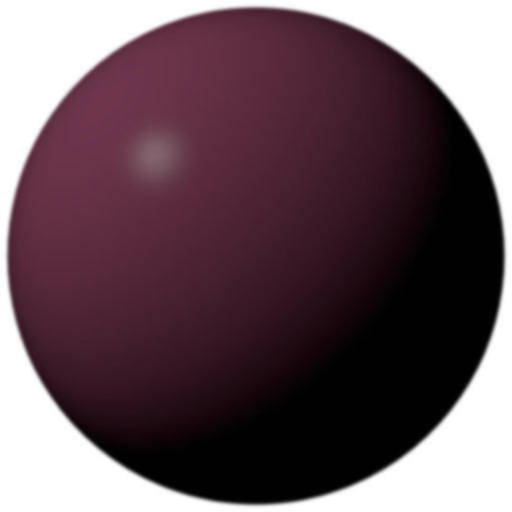} & \includegraphics[width=0.28\linewidth]{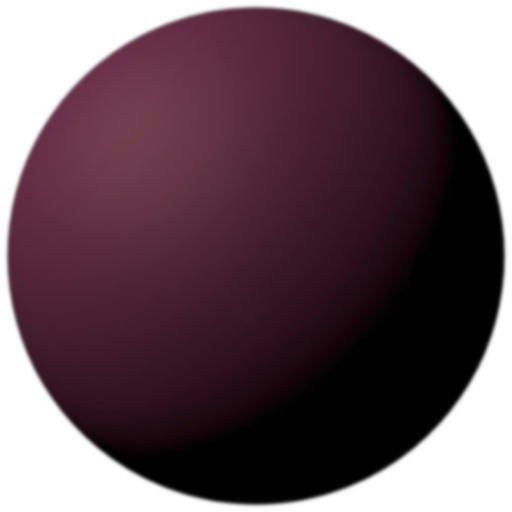} & \includegraphics[width=0.28\linewidth]{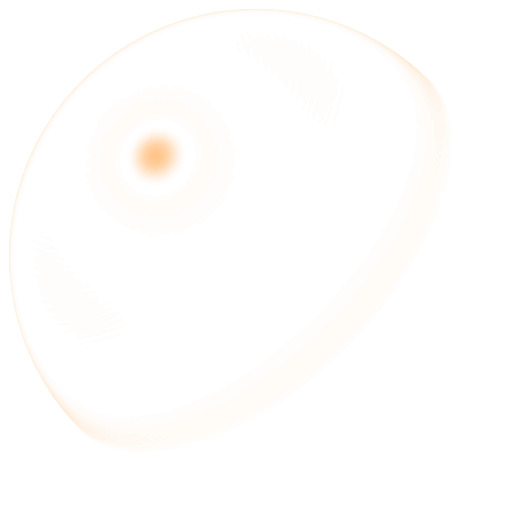} & \\
\end{tabular}
\end{center}
\caption{\label{fig:1456and2410}%
  Source model (\emph{left}), remapped target (\emph{center}) and SSIM error (\emph{right}), corresponding to $IOR=1.3$ (\emph{top}) and $IOR=1.34$ (\emph{bottom}) in Figure~\ref{fig:as2ward-plastics1stage-diffmat3-finalplot-onlyalpha001}.}
\end{figure}

In this case, the instability signals a change of behaviour in the remapping process.
In Figure~\ref{fig:1456and2410} we show renderings that correspond to the points at both
sides of the jump in the curve of Figure~\ref{fig:as2ward-plastics1stage-diffmat3-finalplot-onlyalpha001}.
In one case the remapping is working correctly, and we obtain a similar appearance in
both models. In the other we observe that the optimisation arrives at a local minimum
and the remapping is unable to recover the characteristic highlight from the source.
\setlength{\tabcolsep}{.16em}

\begin{figure}[htb]
  \centering
  \includegraphics[width=\linewidth]{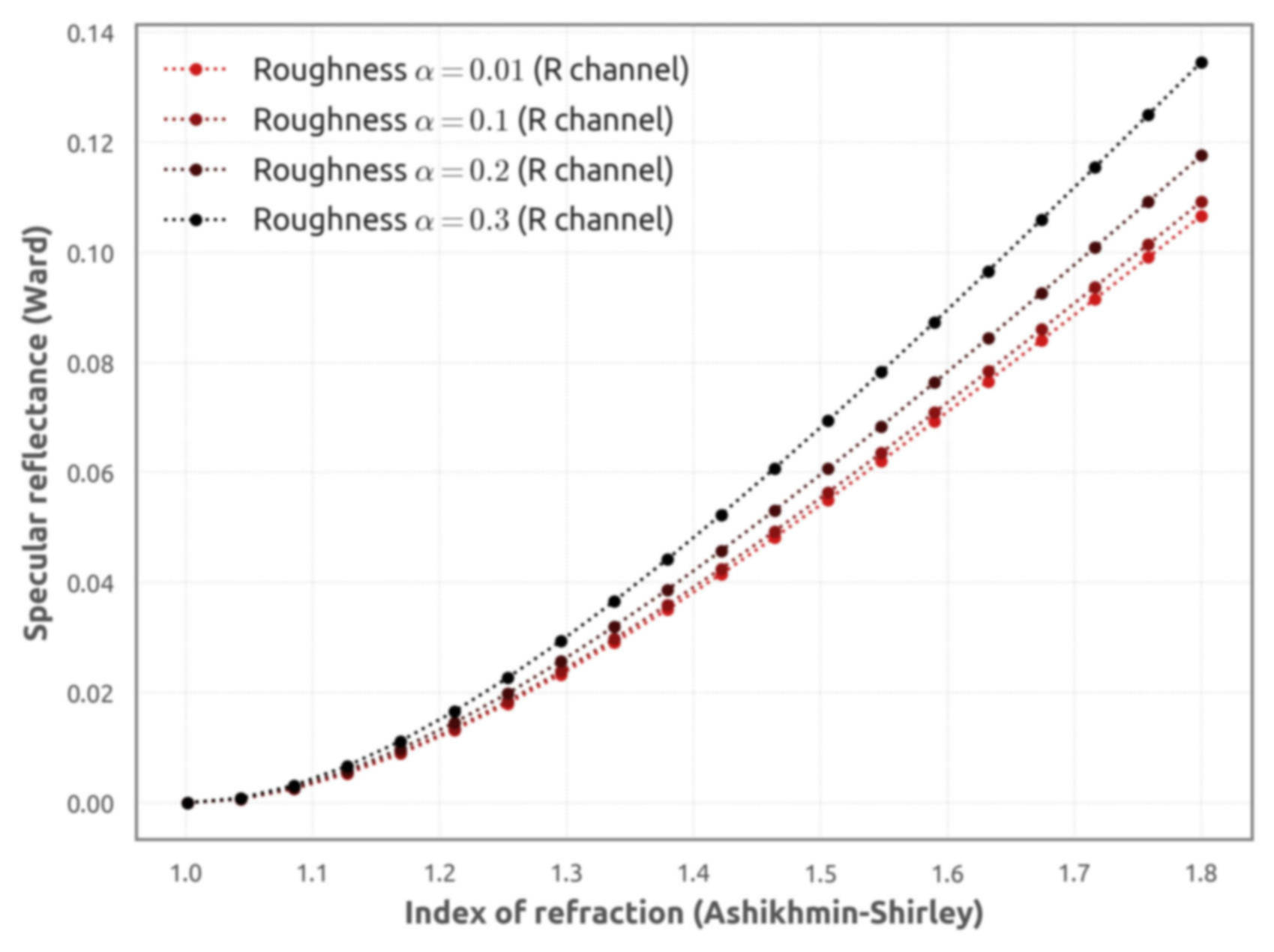}
  \caption{\label{fig:2stage-results}%
    Two-stage remapping of conductors from Ashikhmin-Shirley to Ward. Specular reflectance vs Fresnel coefficient $F_0$ for multiple values of roughness.}
\end{figure}

\begin{figure}[htb]
  \centering
  \includegraphics[width=\linewidth]{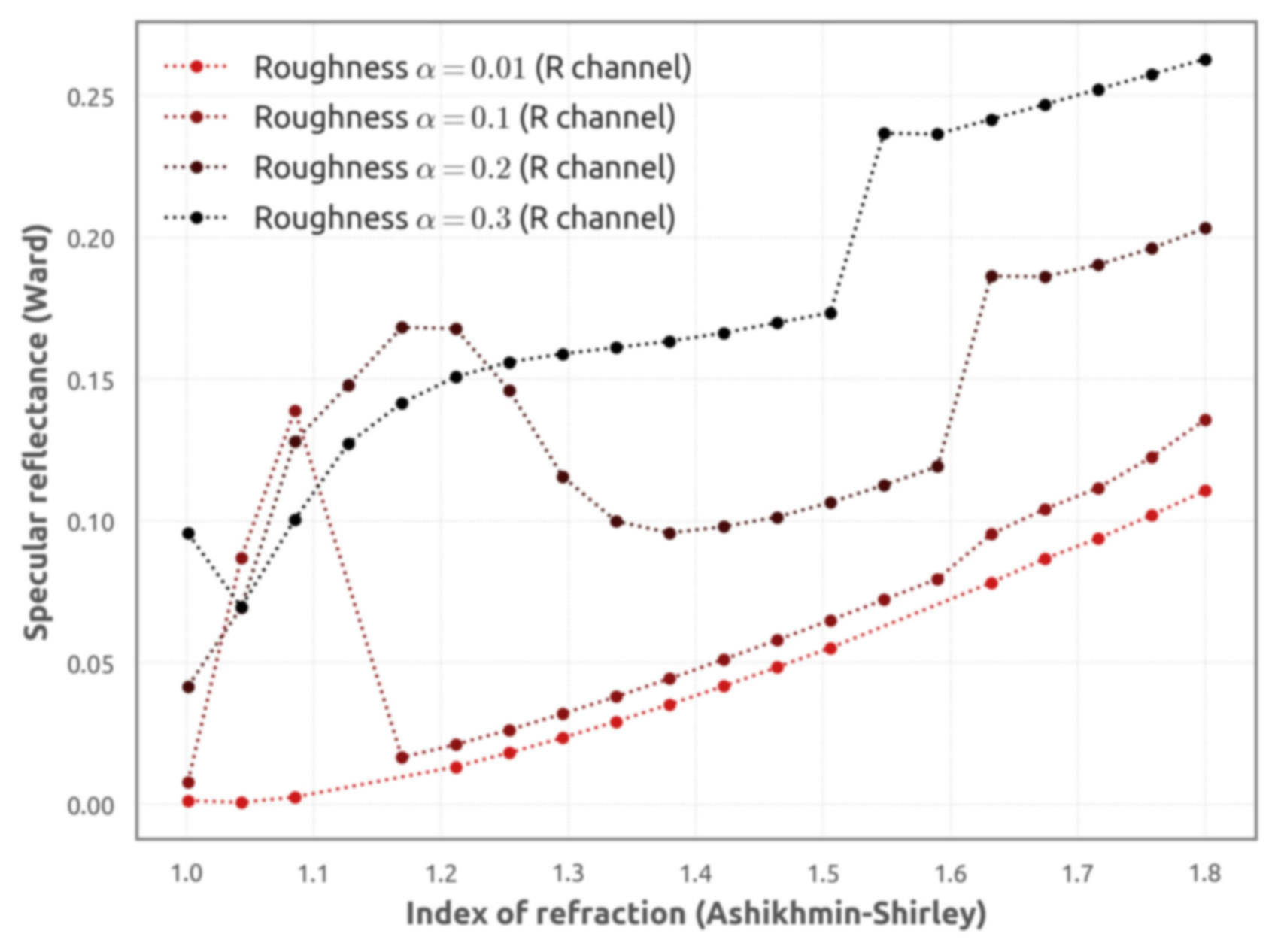}
  \caption{\label{fig:3stage-results}%
    Three-stage remapping of conductors from Ashikhmin-Shirley to Ward. Specular reflectance vs Fresnel coefficient $F_0$ for multiple values of roughness.}
\end{figure}

Figures~\ref{fig:2stage-results} and~\ref{fig:3stage-results} show the results
of the two- and three-stage approaches, developed to improve the stability of the
remapping process and illustrated on the diagrams of
Figures~\ref{fig:scheme-remapping-2stage} and~\ref{fig:scheme-remapping-3stage}.
The two-stage approach effectively recovers a smooth relationship between the parameters,
by avoiding the coupling between the diffuse and specular terms. With the additional
optimisation step of the three-stage approach, in some cases we were able to slightly
reduce the optimisation error with respect to the two-stage approach, but unfortunately
the coupling between diffuse and specular terms still causes several instabilities which
make this second approach unreliable. In summary, in order to generate a robust remapping
we need to avoid the coupling of the diffuse and specular components, by remapping each
term independently (two-stage method).

\section{Remapping of Spatially-Varying Materials}
\label{sec:svbrdf}

\begin{figure}[htb]
  \centering
  \includegraphics[height=5cm]{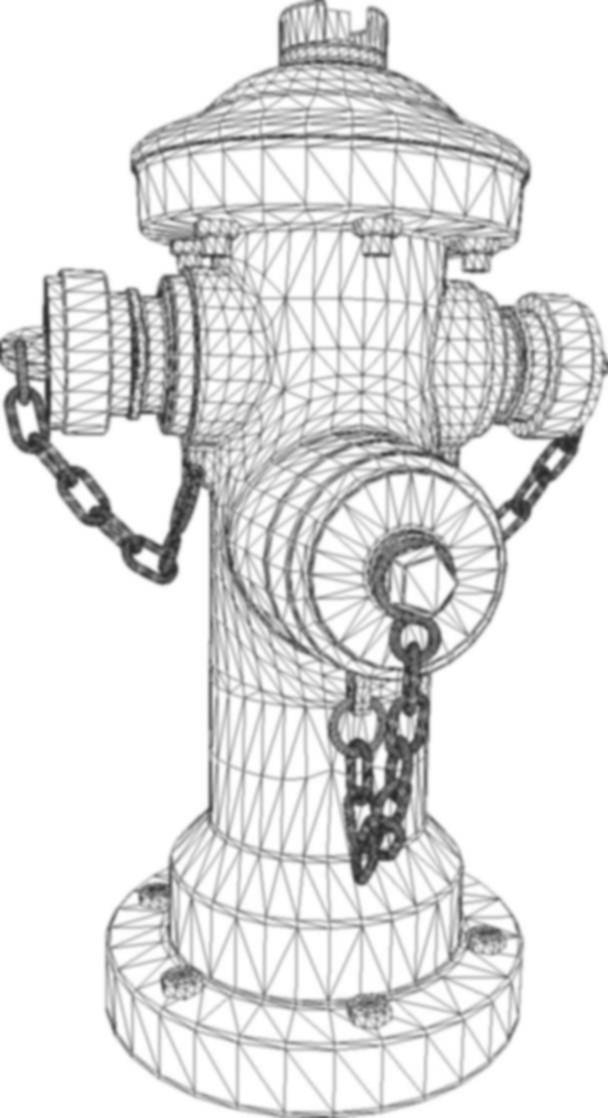}~\hspace{1em}~\includegraphics[height=5cm]{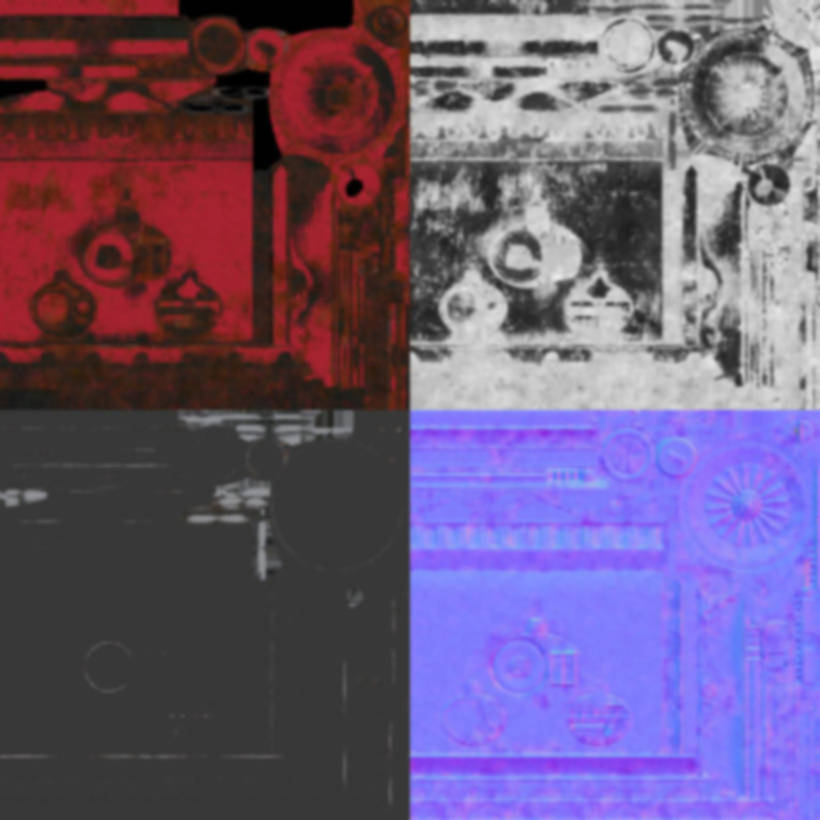}
  \caption{\label{fig:abstract-image1}%
    Decomposition of 3D asset into a low-resolution geometry and
    four texture maps describing the spatially-varying parameters
    of the material (diffuse reflectance, roughness, specular reflectance
    and normals).}
\end{figure}

Spatially-varying materials are commonly defined using texture maps
that provide the value of each model parameter across the surface. In
Figure~\ref{fig:abstract-image1} we display the decomposition of an
SVBRDF asset into texture maps describing four different parameters
involved in the shading process (namely diffuse reflectance, specular
roughness, specular reflectance, and surface normals).\\

The remapping of the corresponding spatially-varying material to 
a different BRDF model or renderer requires the remapping of each individual 
texel, which can be performed using one of the schemes for uniform BRDF 
remapping from the previous section. However, the optimisation required by the 
remapping of a single material usually takes a few minutes, which is acceptable
for uniform BRDFs but intractable for high-resolution texture maps with 
hundred-thousands of texels.

To solve this problem we employ a regression scheme that utilises the 
data from the remapping of uniform materials to learn the relationship between 
parameters in both BRDF models involved. Through parameter sweep over the source 
model we generate a database of uniform material parameters and their remapped 
counter-parts in the target model, which is used as input for a regression scheme.
Thus we are able to generate a reduced representation of the transformation between
the two models, which can then be evaluated efficiently without the need for further
optimisation.

\subsection{Parametric regression scheme}
\label{sec:parametric-scheme}

\begin{figure}[htb]
  \centering
  \includegraphics[width=\linewidth]{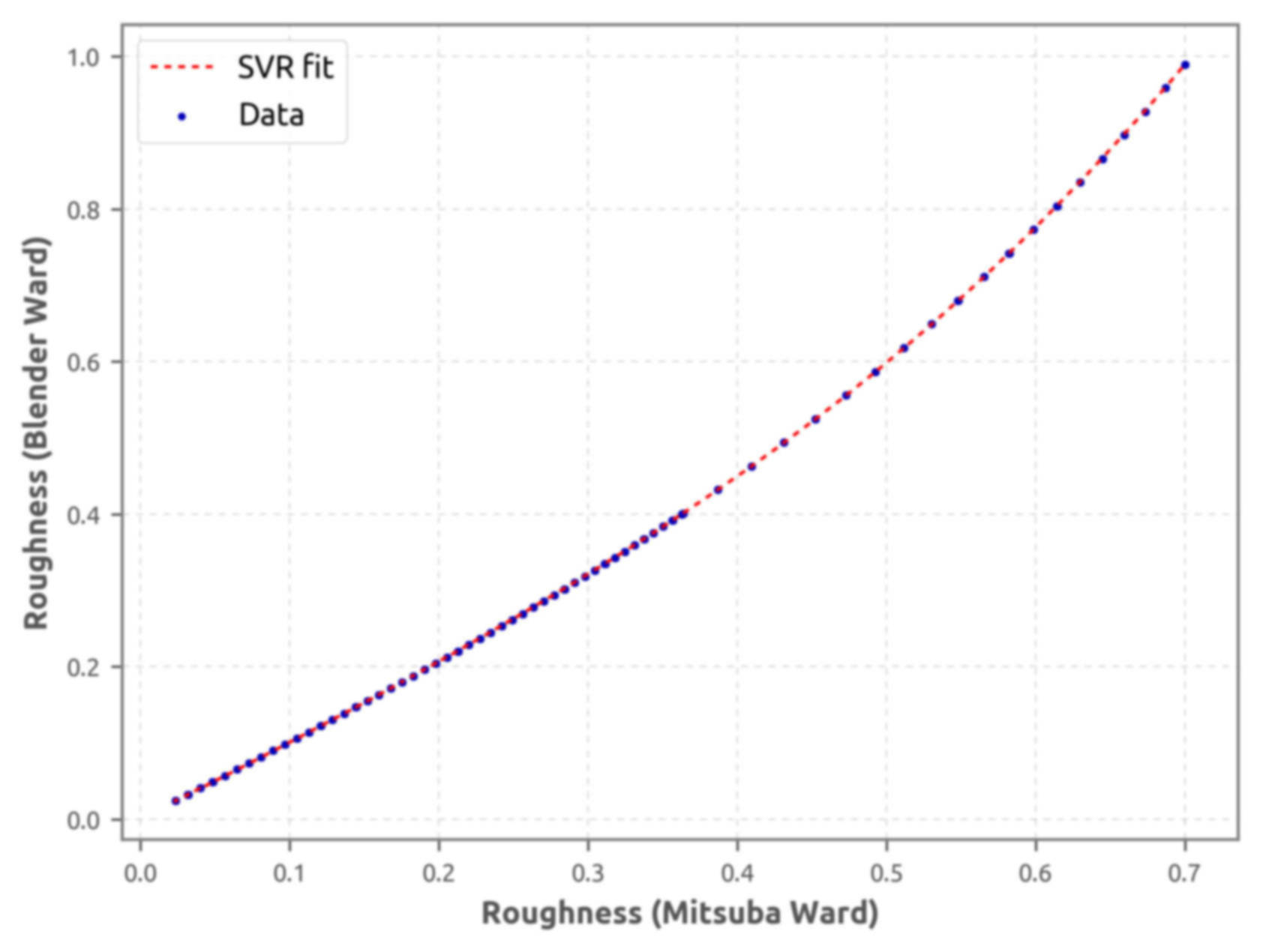}
  \includegraphics[width=\linewidth]{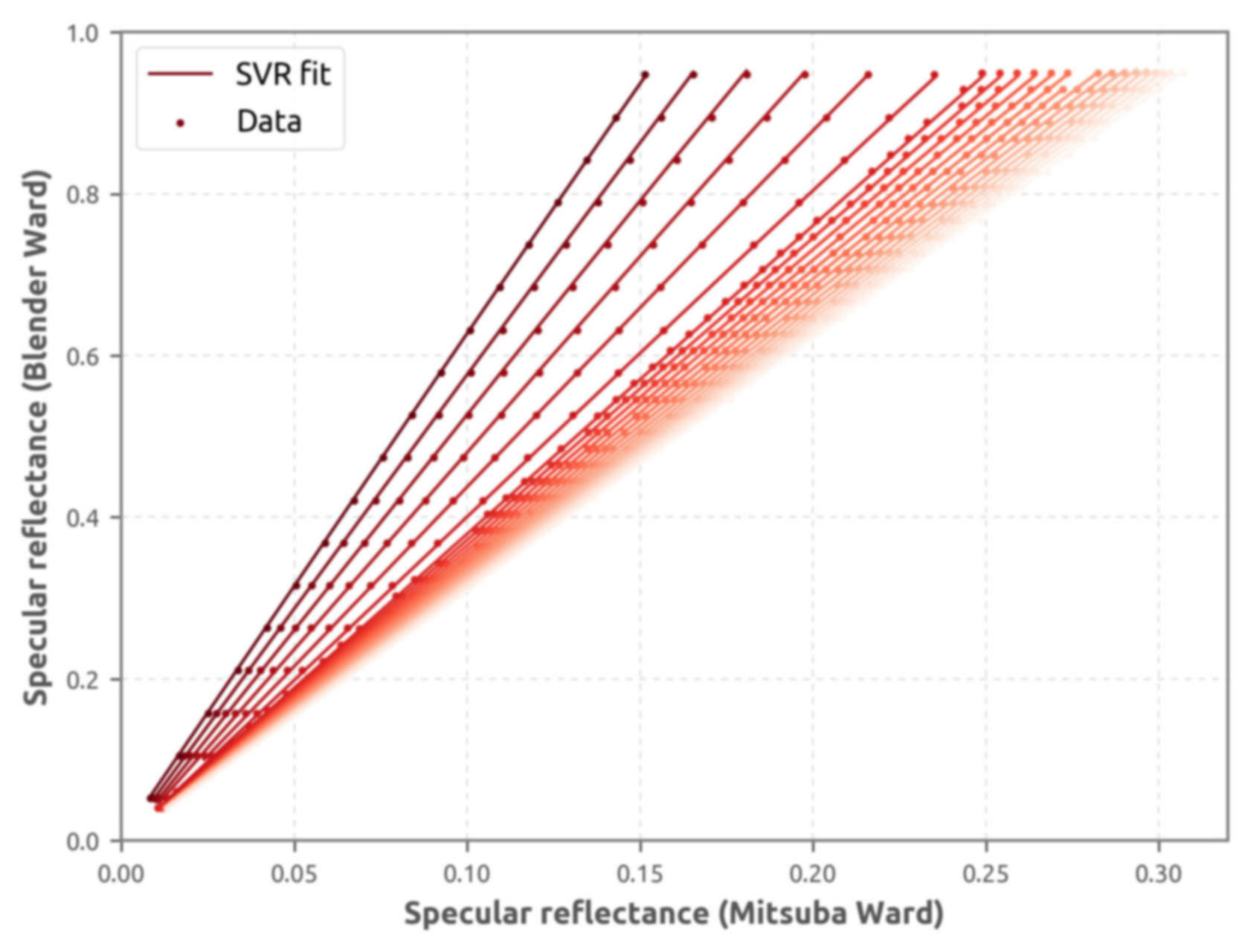}\\
  \includegraphics[width=0.6\linewidth]{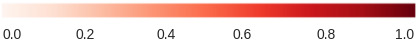}
  \caption{\label{fig:blender-ward_vs_mitsuba-ward_plots}%
     Remapping from Mitsuba-Ward to Blender internal-Ward, with data points and regression.
     \emph{Top:} Remapping of roughness. The non-linear mapping shows that the two
     variants of Ward are distinct.
     \emph{Bottom:} Remapping of specular reflectance with color indicating roughness.
     The transformation follows different mappings depending on the roughness parameter.}
\end{figure}

Figure~\ref{fig:blender-ward_vs_mitsuba-ward_plots} shows an example transformation
that remaps between two implementations of Ward: from Mitsuba to Blender internal renderer.
Due to established variations of well-known BRDF models, but also in order to address
user expectations, such as energy preservation under changes of roughness, individual
renderer implementations may take different design decisions, leading to differences
between models despite having the same name. 

The top plot shows a non-linear
relationship between the two ``roughness'' parameters, illustrating a difference in
implementations of Ward that cannot be explained by a simple rescaling of variables.
The bottom plot offers a different set of
cross-sections through the same remapping of parameter spaces: for a wide range of
roughness values, we show how varying specular reflectance values of Mitsuba's Ward map
to Blender's Ward model parameters. In contrast with the one-to-one correspondence between
parameters found in the top plot, here the transformation traces a different
mapping depending on the roughness parameter, signaling a complex relationship between
multiple parameters in the models which would be hard to recover by a manual mapping of
model parameters.

The mapping of BRDF parameters of Figure~\ref{fig:blender-ward_vs_mitsuba-ward_plots} can be
approximated by complex learning methods such as Support Vector Regression or
Neural Networks. However, upon further scrutiny of the functional shape of the transformation,
which presents common properties along different BRDF model pairs, we are able to formulate
a parametric function able which capitalizes on these properties and is able to model the
behaviour of the transformation with only a few parameters. Below we analyse the properties
of the transformations and formulate the parametric approach for regression.
In Section~\ref{sec:results:svr} we will compare this approach with Support Vector Regression and show that,
in addition to being easier to train, it presents a much better extrapolation of BRDF
parameters outside of the region sampled by the training set.

In the top plot in Figure~\ref{fig:blender-ward_vs_mitsuba-ward_plots}  we observe that the
remapping of the roughness parameter depends only on the roughness. This behaviour is
easy to model with a simple univariate polynomial fit of low-degree ($\leq 4$).
In contrast, the remapping of the specular parameter (bottom) depends on both the
roughness and the specular parts, presenting a more complex functional for the regression.
However, the complexity
of the regression can be drastically reduced by observing that for a fixed value of roughness,
the relationship between specular parameters results in a very low-degree polynomial.
In particular, when the specular parameters are linearly related to the intensity of the
BRDF lobe, the specular transformation results in straight lines. (in the case of the IOR,
as seen in Figure~\ref{fig:2stage-results}, the non-linear change of variable of equation
\ref{eq:f0} can be used to recover the specular reflectance). Thus, our parametric model
for the specular transformation results in:
\begin{equation}
\text{s}_2(\text{s}_1, \alpha_1) = k(\alpha_1) \cdot \text{s}_1
\label{eq:parametric-function}
\end{equation}
where $s$ and $\alpha$ refer to specular and roughness parameters, and the subindices $1$
and $2$ indicate original, and remapped, respectively.
With this parametric model, all that we have to do is compute the slope of the curves for
each value of roughness, and then use this data to fit the non-linear relationship
$k(\alpha_1)$ between roughness and slope. This can be done with a univariate
non-linear fit with few coefficients $c$, such as:
\begin{equation}
k(\alpha) = c_0 + c_1 e^{-c_2 \alpha} + c_3 e^{-c_4 \alpha^2}
\label{eq:kofalpha}
\end{equation}

\subsection{Results of Cross-Renderer SVBRDF Remapping}
\label{sec:results-svbrdf-remapping}

In line with the results from previous sections, we now proceed by basing all BRDF
transformations on the robust two-stage remapping that operates on diffuse and specular
components independently.
For each pair of BRDF models of interest, we generate a database of mutually
corresponding uniform material parameters and use this as input for a regression scheme
that learns the mapping between models. This generates a reduced
representation of the transformation, which can be evaluated swiftly in any part of the
parameter space and without further optimisation or interaction with the renderer.
Note that, even though our uniform remapping implementation requires the target BRDF
to be defined within Mitsuba, we can reverse the role of source and target
in the regression process, thus implementing BRDF remappings in the opposite direction.

\begin{figure}[htb]
  \centering
  \includegraphics[width=.49\linewidth]{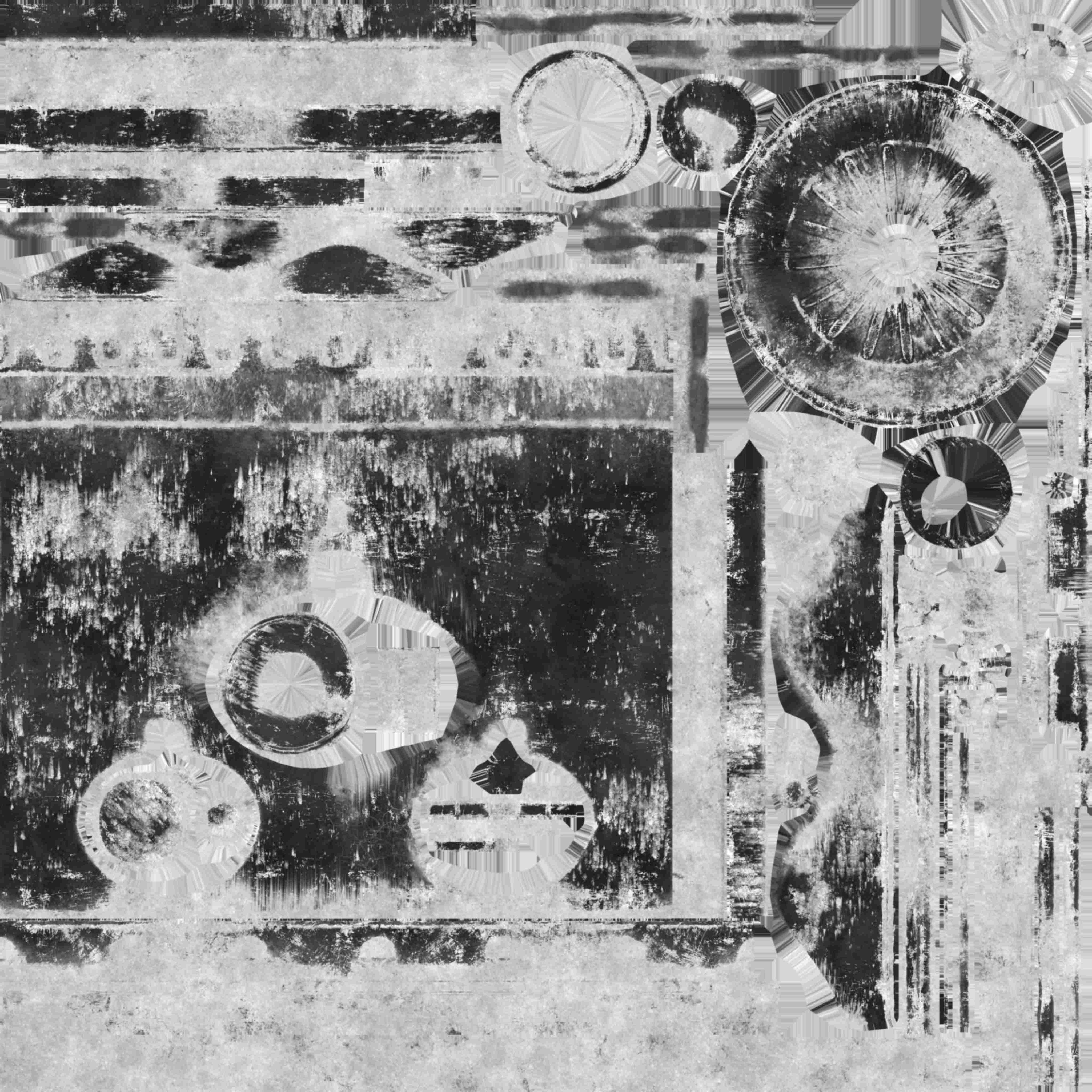}~\includegraphics[width=.49\linewidth]{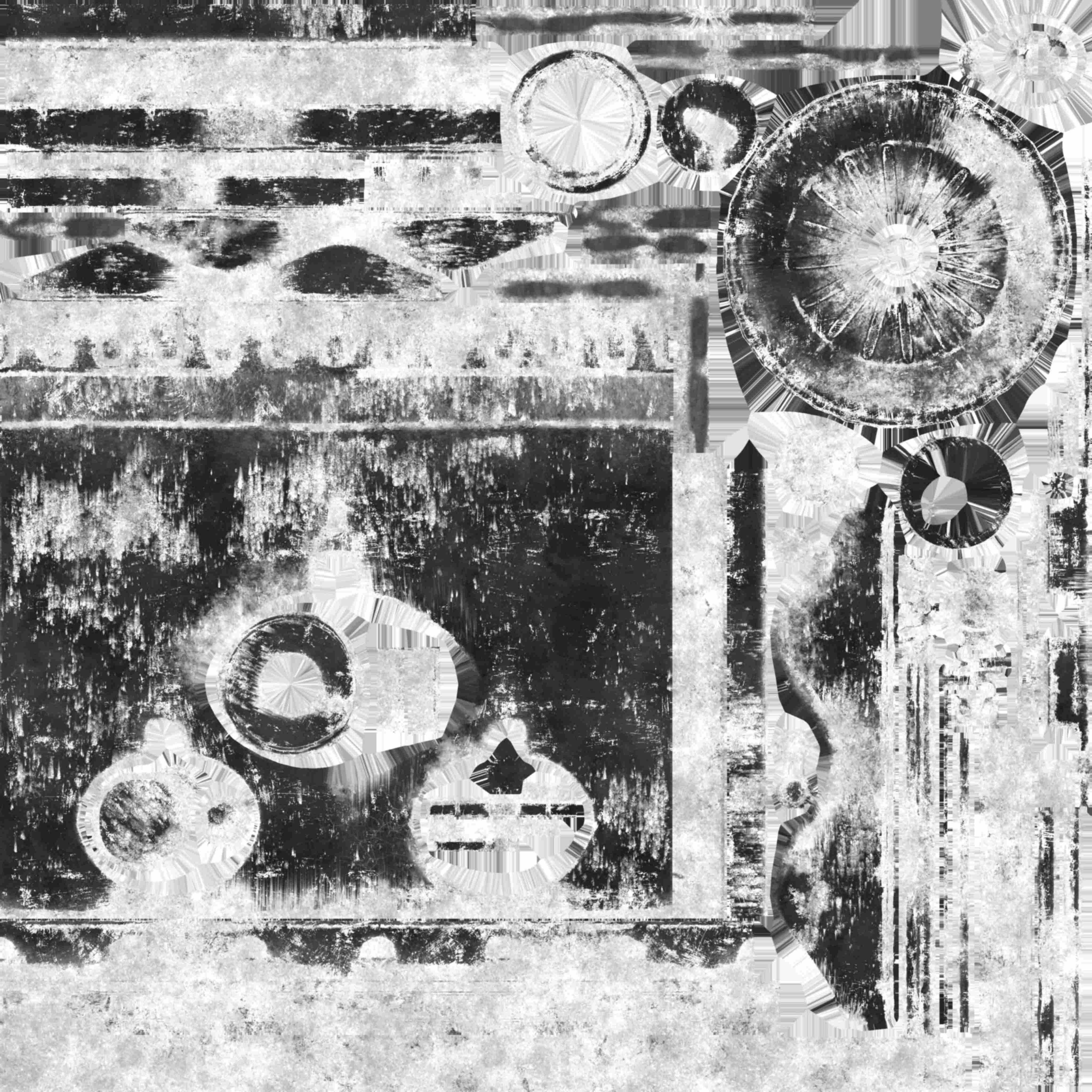}
  \caption{\label{fig:roughness_textures}%
    Roughness texture maps. \emph{Left:} original Mitsuba's Ward. \emph{Right:} remapped Blender's ward.}
\end{figure}

\begin{figure}[htb]
  \centering
  \includegraphics[width=.49\linewidth]{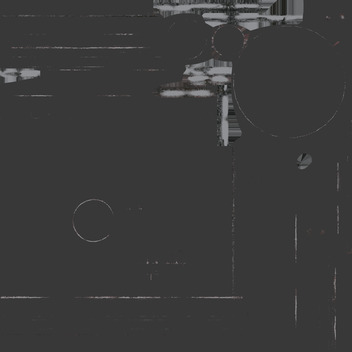}~\includegraphics[width=.49\linewidth]{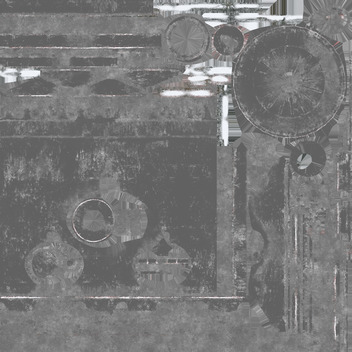}
  \caption{\label{fig:specular_textures}%
    Specular texture maps. \emph{Left:} original Mitsuba's Ward. \emph{Right:} remapped Blender's ward.\vspace*{-1.5ex}}
\end{figure}

In Figure~\ref{fig:blender-ward_vs_mitsuba-ward_plots} we analysed the remapping between two
implementations of Ward: from Mitsuba to Blender internal renderer. Once the compact
representation of the transformation is generated, it can be used to perform the remapping
of every texel in the parameter maps of an SVBRDF. In Figures~\ref{fig:roughness_textures}
and~\ref{fig:specular_textures} we display the roughness and specular maps for
Mitsuba-Ward and their corresponding remapped versions in Blender internal-Ward.

As previously seen, the transformation of the roughness
(Figure~\ref{fig:blender-ward_vs_mitsuba-ward_plots} top) depends only on the original
roughness, and thus the remapping of a roughness texture map is essentially a tonemapping
operation, as observed in Figure~\ref{fig:roughness_textures}.
In contrast the remapping of the specular parameter
(Figure~\ref{fig:blender-ward_vs_mitsuba-ward_plots} bottom) depends on both the roughness
and the specular parts, which produces a remapped specular map that inherits details of
the roughness map (Figure~\ref{fig:specular_textures}).

\begin{figure}[htb]
  \centering
  \includegraphics[width=.49\linewidth]{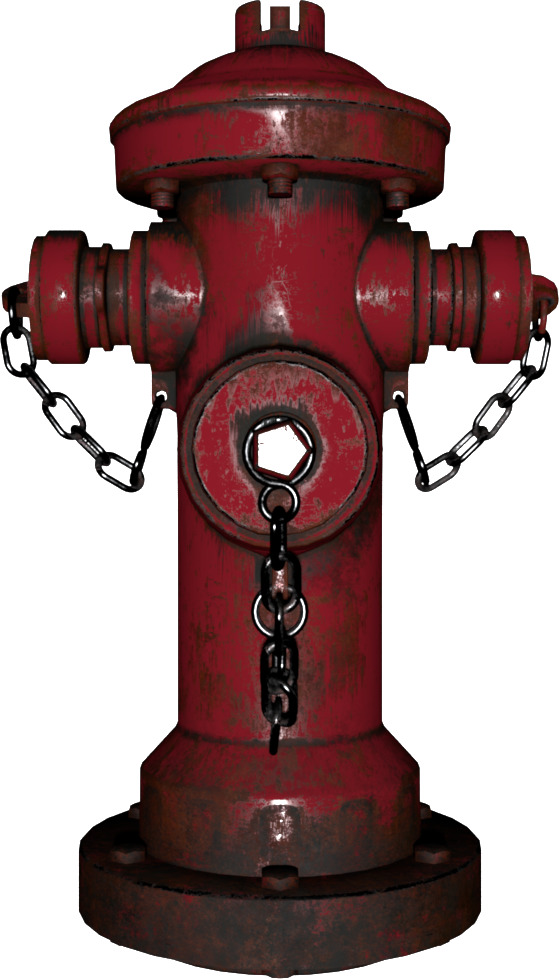}
  \includegraphics[width=.49\linewidth]{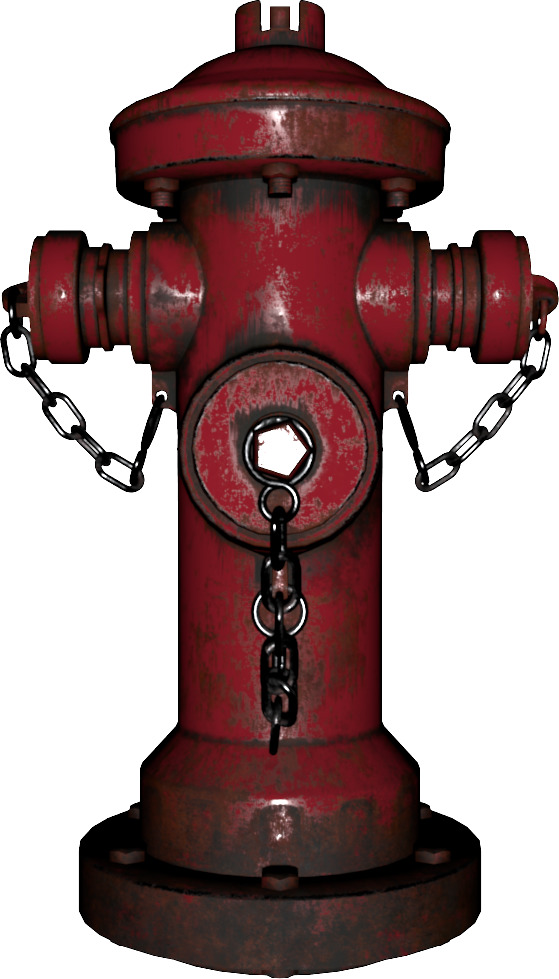}\\[1ex]
  \includegraphics[height=3.6cm]{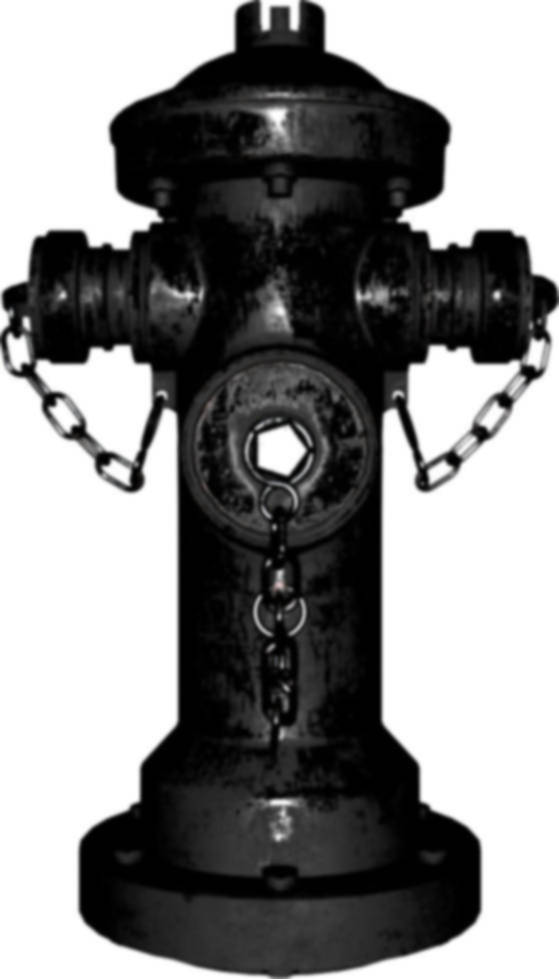}
  \includegraphics[height=3.6cm]{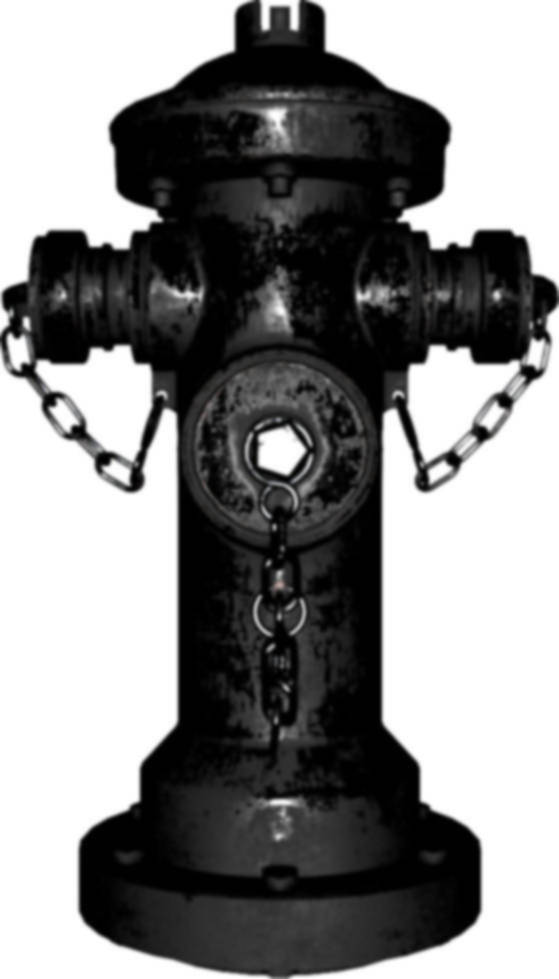}
  \includegraphics[height=3.6cm]{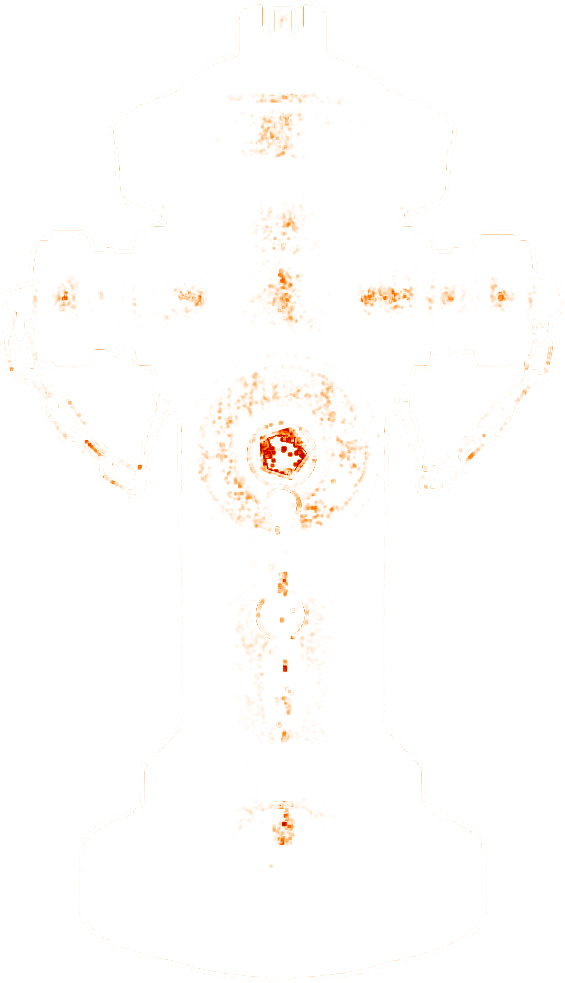}
  \enskip\includegraphics[height=3.6cm]{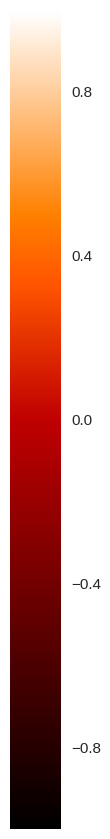}

  \caption{\label{fig:blender-ward_vs_mitsuba-ward_renderings}%
     SVBRDF remapping from Mitsuba's default Ward implementation to Ward within Blender's
     internal renderer. \emph{Top:} corresponding renderings using Mitsuba \emph{(left)}
     and Blender \emph{(right)}. \emph{Bottom:} specular-only renderings with SSIM error
     image determined in linear HDR ($SSIM = 1$ indicates absolute similarity).%
     }
\end{figure}

Figure~\ref{fig:blender-ward_vs_mitsuba-ward_renderings} displays the rendering of a 3D asset
using the texture maps from Figures~\ref{fig:roughness_textures} and \ref{fig:specular_textures}.
The corresponding
renderings show the efficacy of the remapping: faint visual differences are limited
to surface parts that face both camera and light source.
The bottom row shows specular
in isolation, including an SSIM error image ($SSIM = 1$ indicates absolute similarity);
removing the diffuse term, which is
very similar across the renderers, highlights visual differences further.

\begin{figure}[htb]
	\centering
    \includegraphics[height=4.5cm]{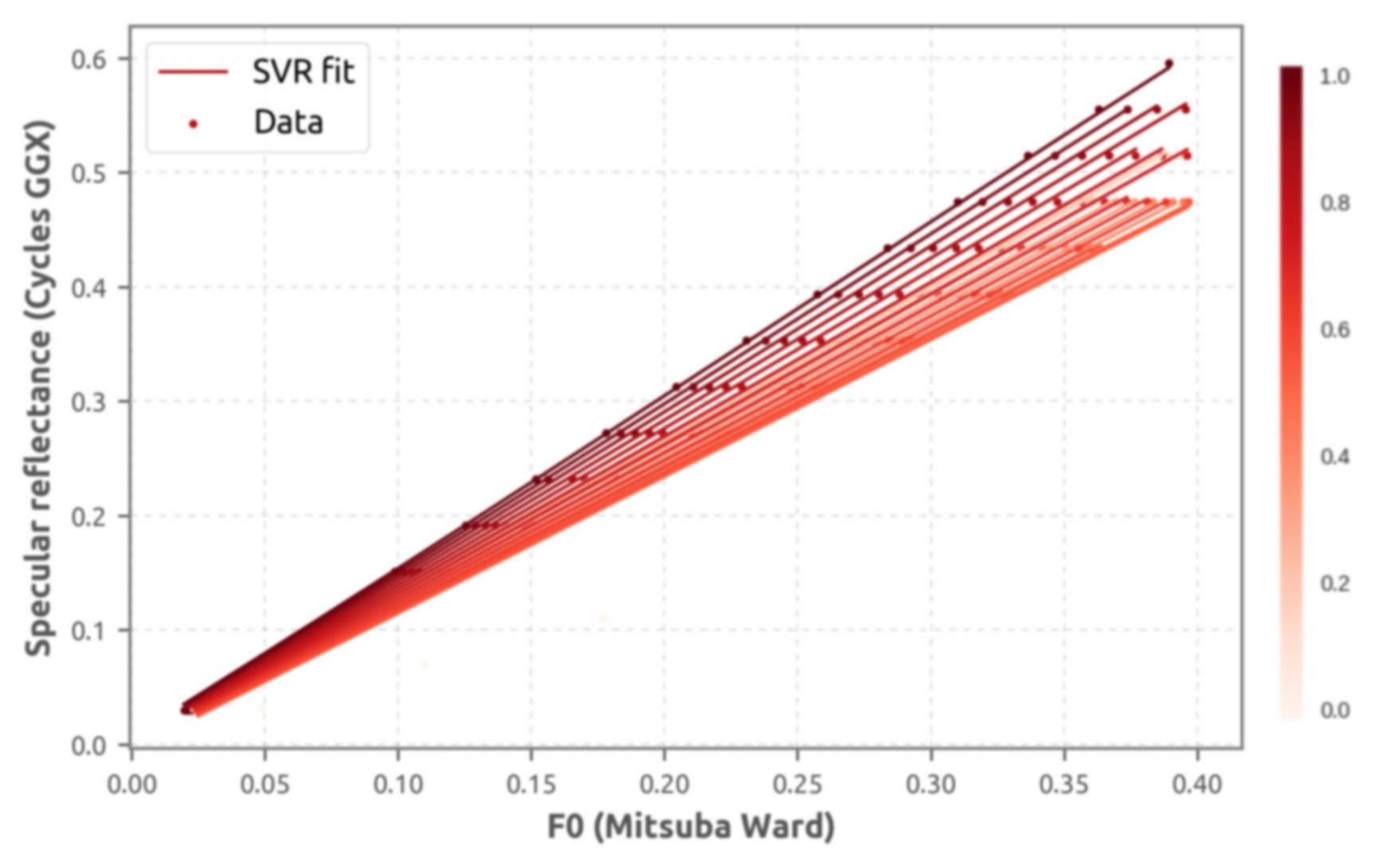}\\
    \hspace{-1em}\includegraphics[height=4.5cm]{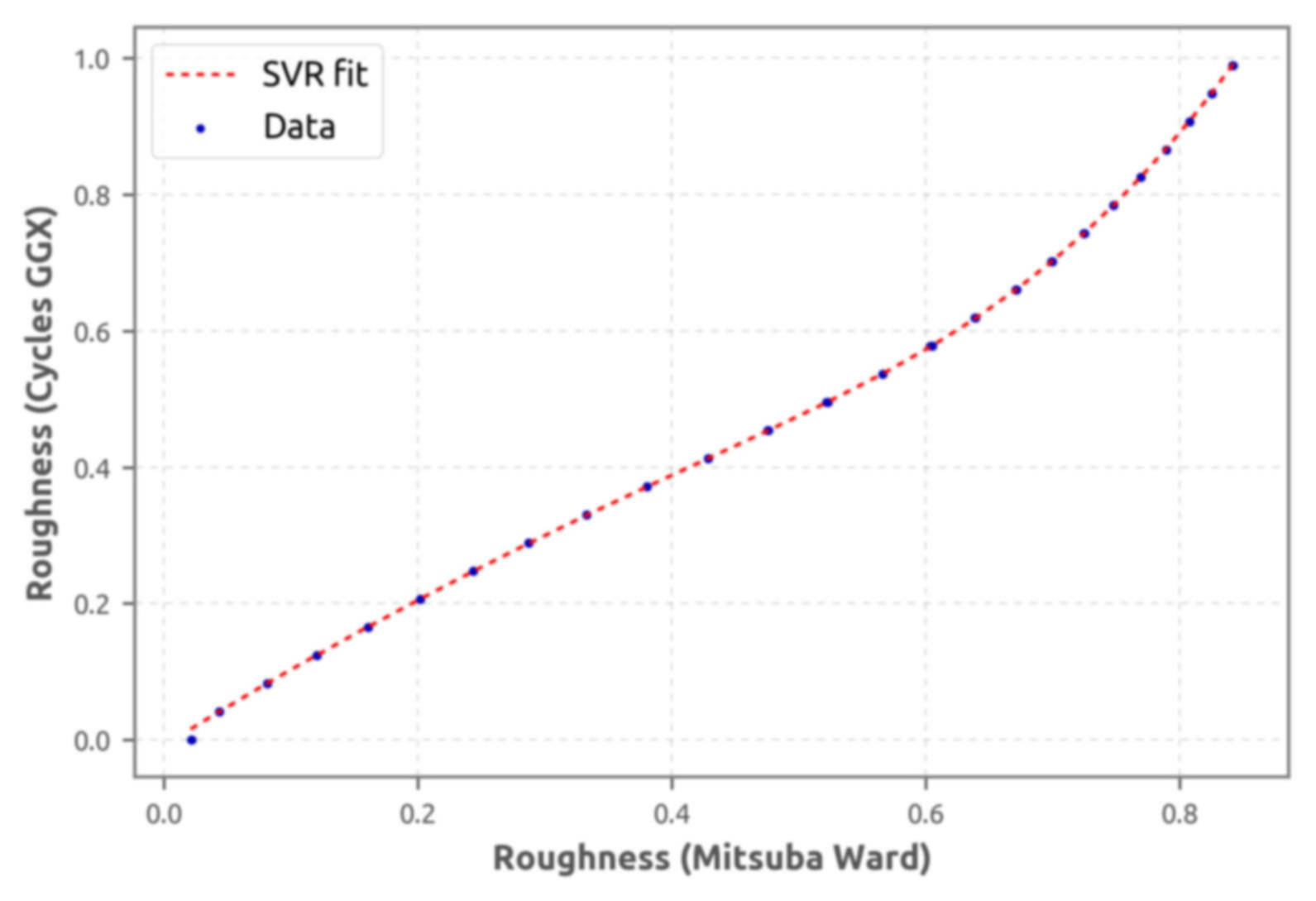}\\

    \vspace{1em}

    \includegraphics[height=4cm]{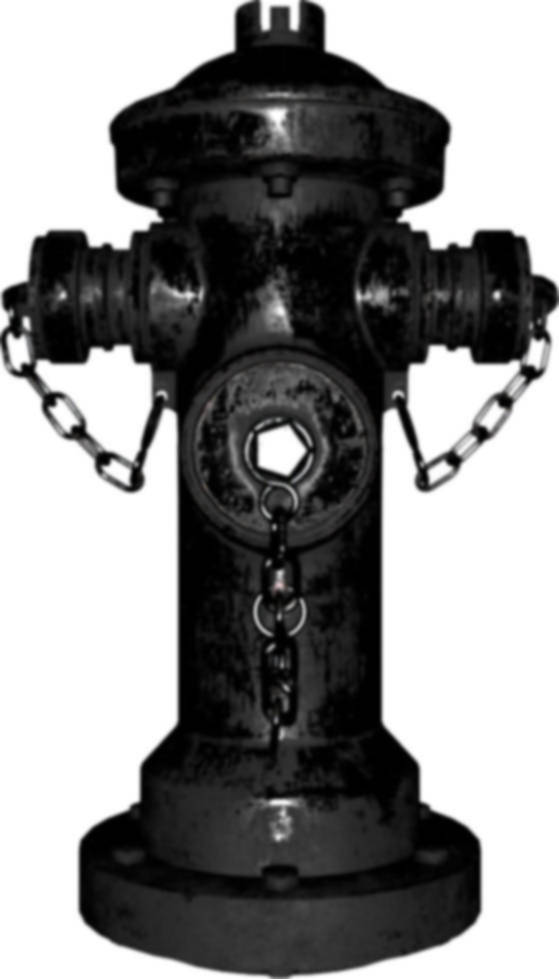}
    \includegraphics[height=4cm]{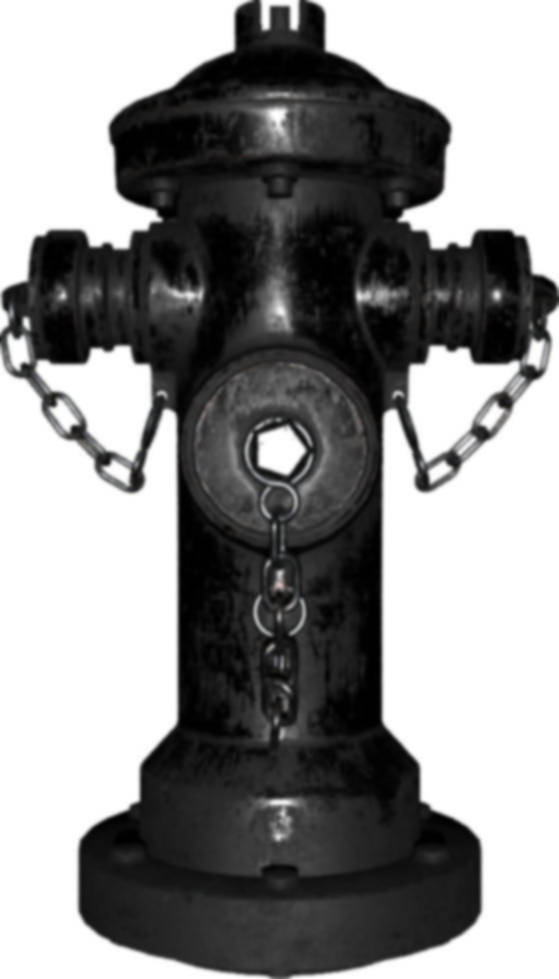}
    \includegraphics[height=4cm]{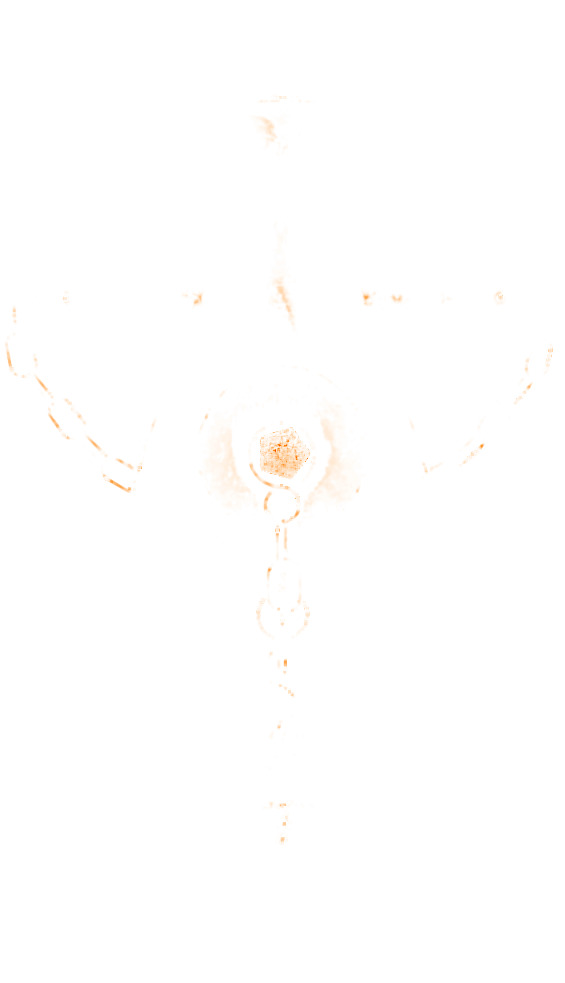}
    \enskip\includegraphics[height=4cm]{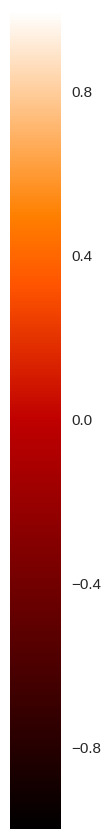}

	\caption{\label{fig:mits-as_to_cycl-ggx}%
    Remapping from Mitsuba Ashikhmin-Shirley to Cycles GGX.
    \emph{Top:} remapping of Fresnel coefficient $F_0$ (left) with color indicating roughness, and
    remapping of roughness (right).
    \emph{Bottom:} specular-only renderings with difference image.
	}
\end{figure}

Figure~\ref{fig:mits-as_to_cycl-ggx} displays a remapping
from Mitsuba Ashikhmin-Shirley shader to Cycles' GGX model. Once again, we show
cross-sectional plots of the parameter remapping function, as well as specular-only
renderings with difference image. This result is of particular interest, as the specular
lobe of GGX significantly differs from traditional microfacet models, such as
Ashikhmin-Shirley. With its heavy tails, the specular term tends to add persistent sheen
to a surface, making it challenging to match the appearance of a model with more compact
reflectance lobes. Considering that, we believe that our remapping preserves the overall
appearance exceptionally well.

\begin{figure}[htb]
	\centering
    \includegraphics[height=4.5cm]{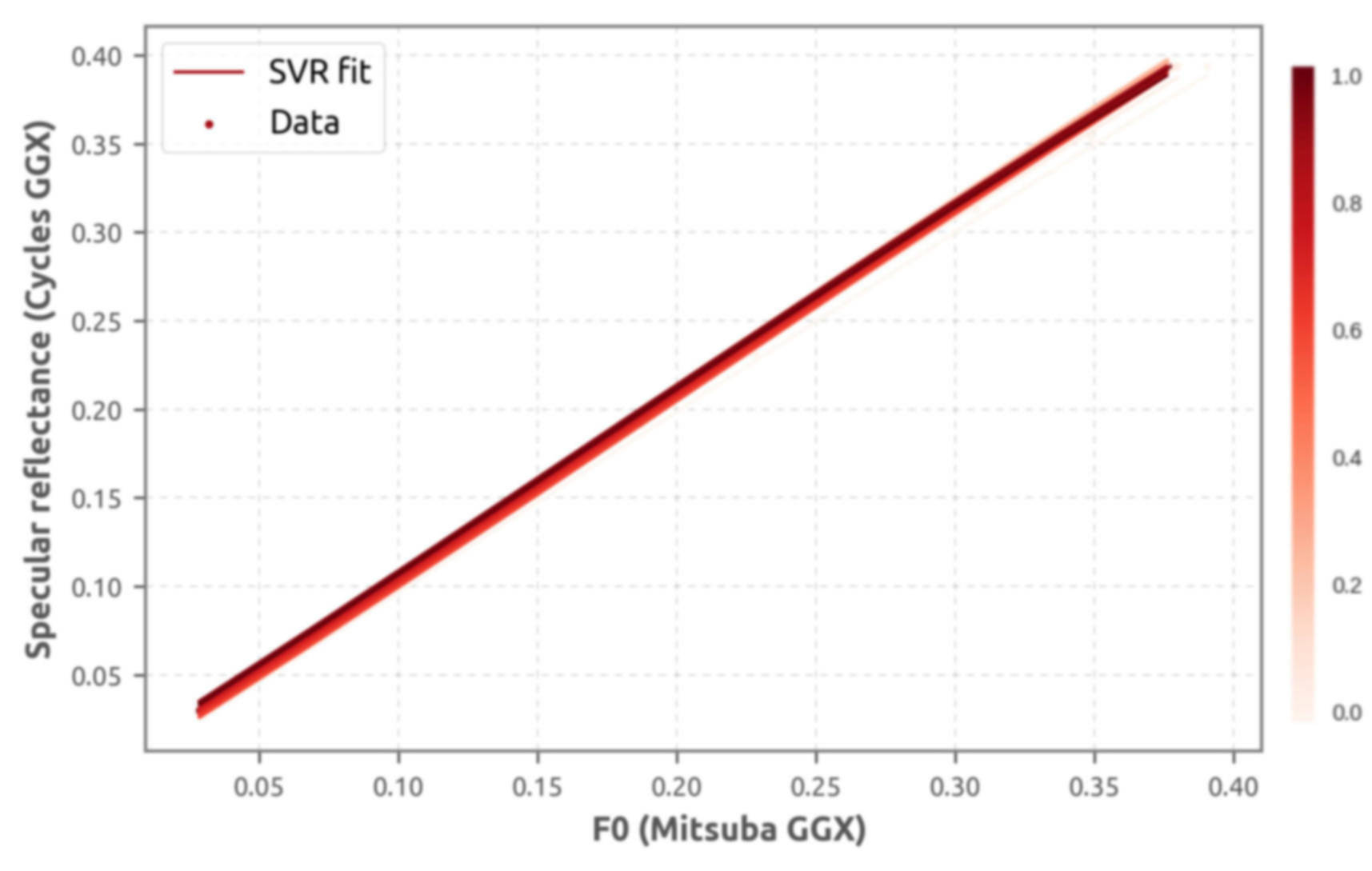}\\
    \hspace{-1em}\includegraphics[height=4.5cm]{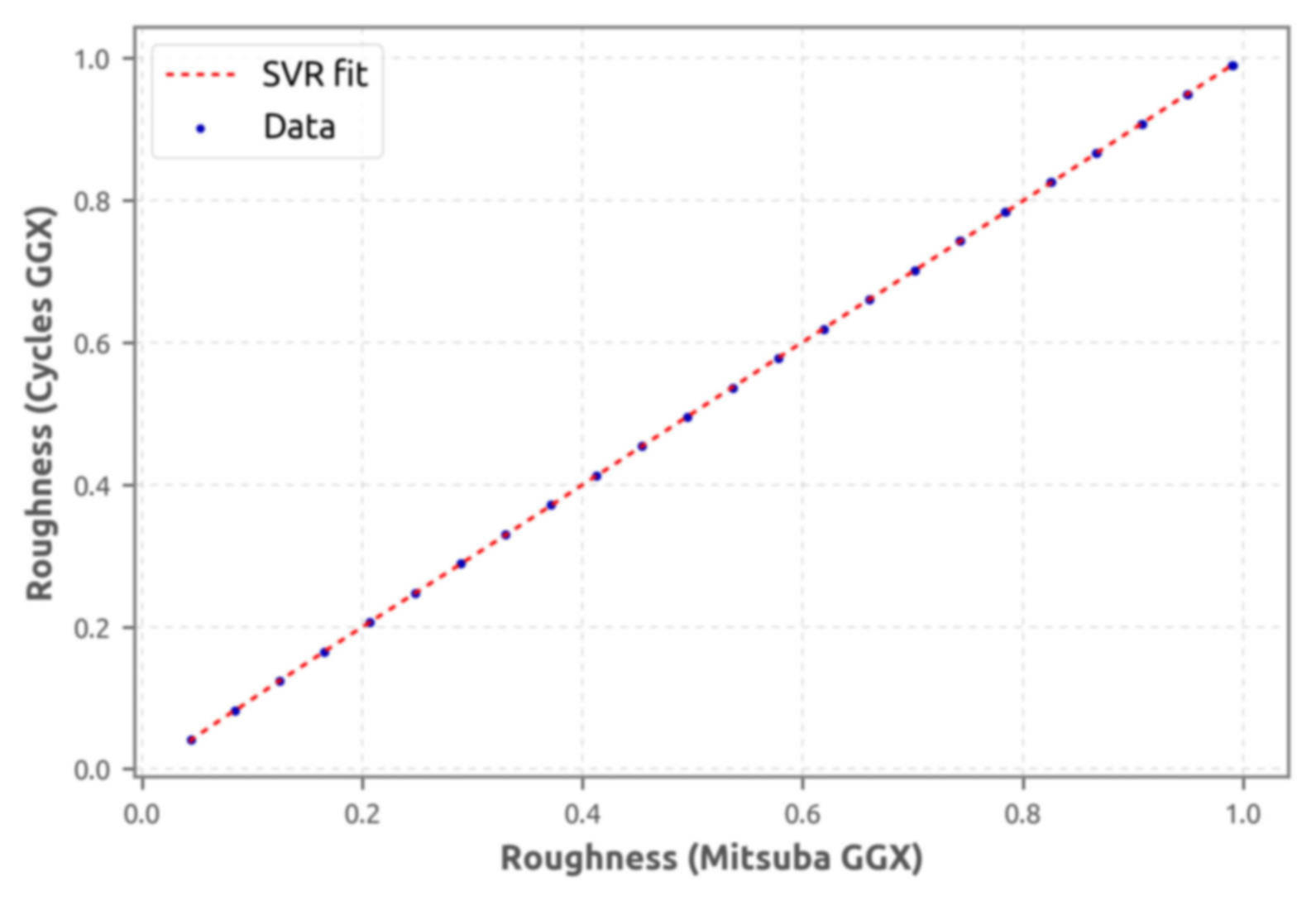}\\

    \vspace{1em}

  \includegraphics[height=4cm]{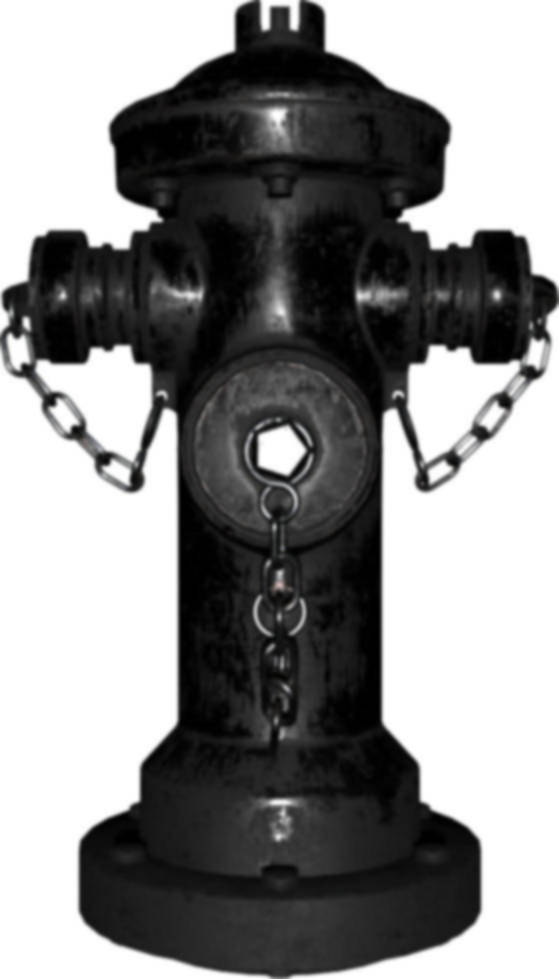}
  \includegraphics[height=4cm]{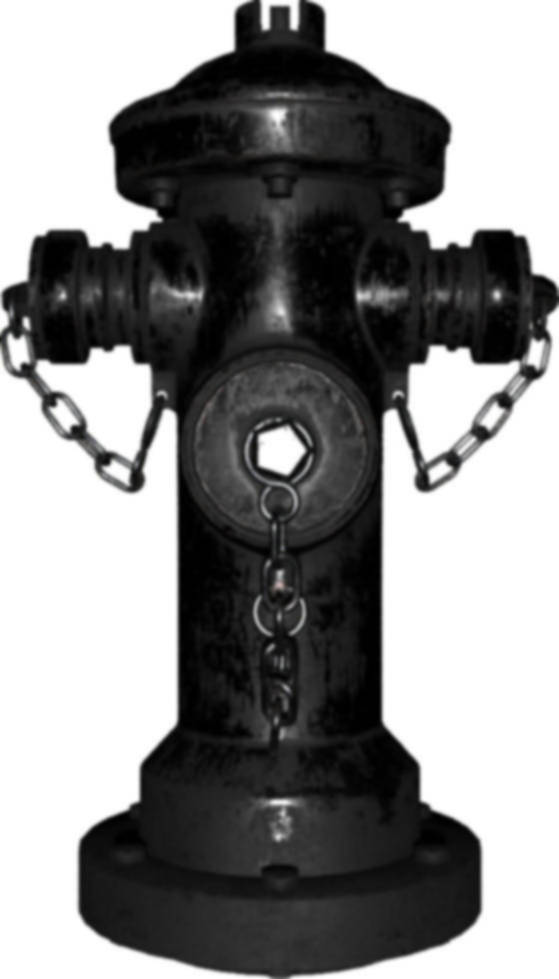}
  \includegraphics[height=4cm]{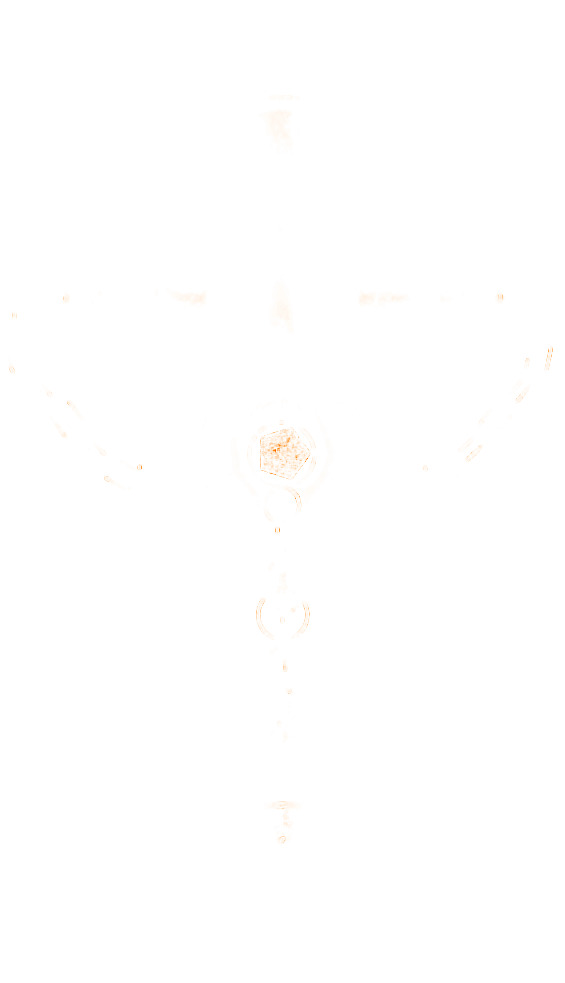}
  \enskip\includegraphics[height=4cm]{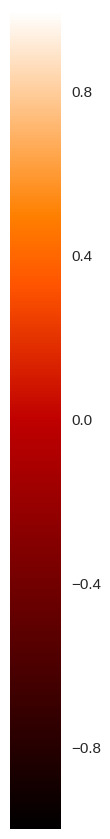}

  \caption{\label{fig:from-mitsuba-ggx-to-cycles-ggx}%
    Remapping from Mitsuba GGX to Cycles GGX.
    \emph{Top:} Remapping of Fresnel coefficient $F_0$ (left) with color indicating roughness, and
    remapping of roughness (right).
    \emph{Bottom:} Specular-only renderings with SSIM error.
  }
\end{figure}

Even within implementations of GGX, however, we observe slight differences between
renderers, as can be seen in Figure~\ref{fig:from-mitsuba-ggx-to-cycles-ggx} where we
display a remapping from Mitsuba-GGX to Blender Cycles-GGX. Here the relationship between
parameters is linear and only shows slight deviations from the identity transformation
for the Fresnel coefficient $F_0$. The simplest explanation for this behaviour is a small difference
in the parameterisation of $F_0$ in each model, although other contributions
can not be discarded (e.g., small differences in the light sources in each renderer).

\begin{figure*}[hbt!]
  \centering
  \includegraphics[height=3.9cm]{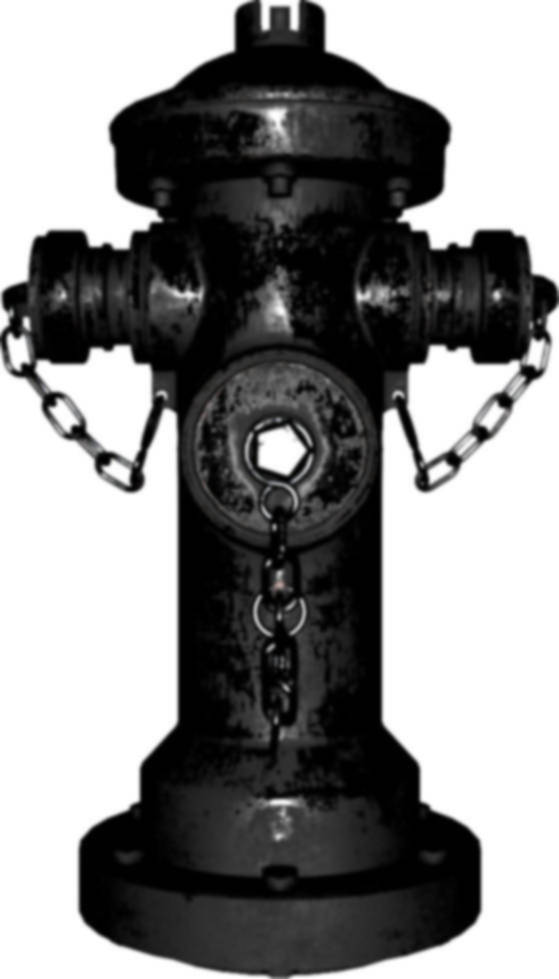}
  \includegraphics[height=3.9cm]{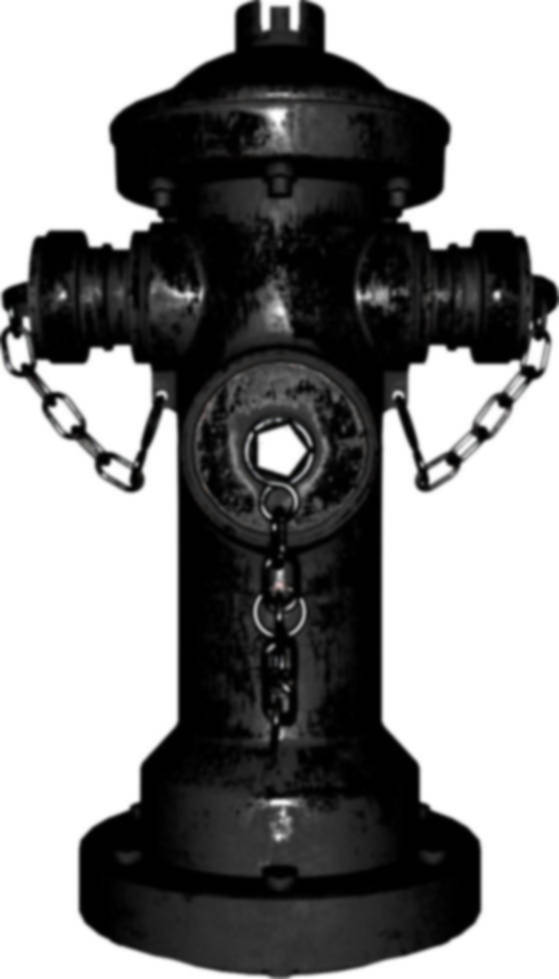}
  \includegraphics[height=3.9cm]{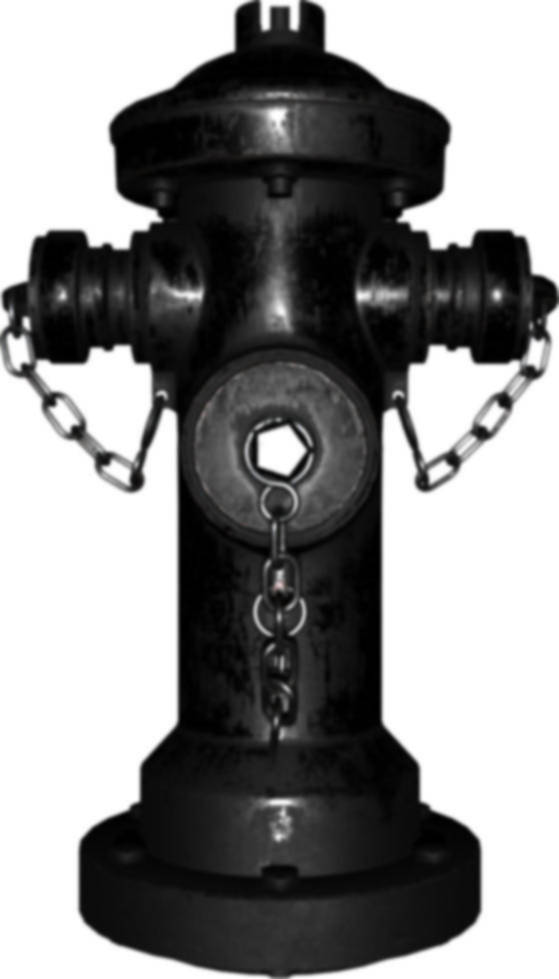}
  \includegraphics[height=3.9cm]{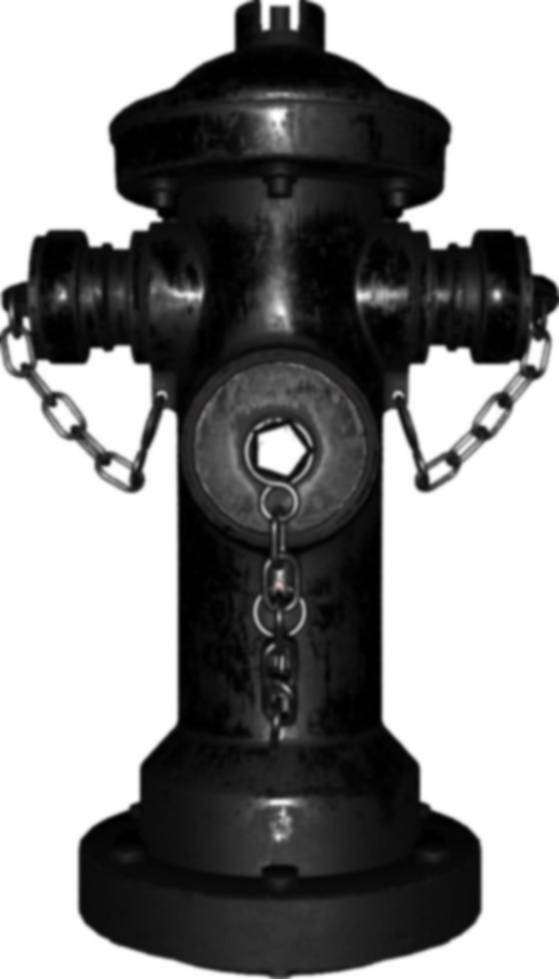}
  \includegraphics[height=3.9cm]{{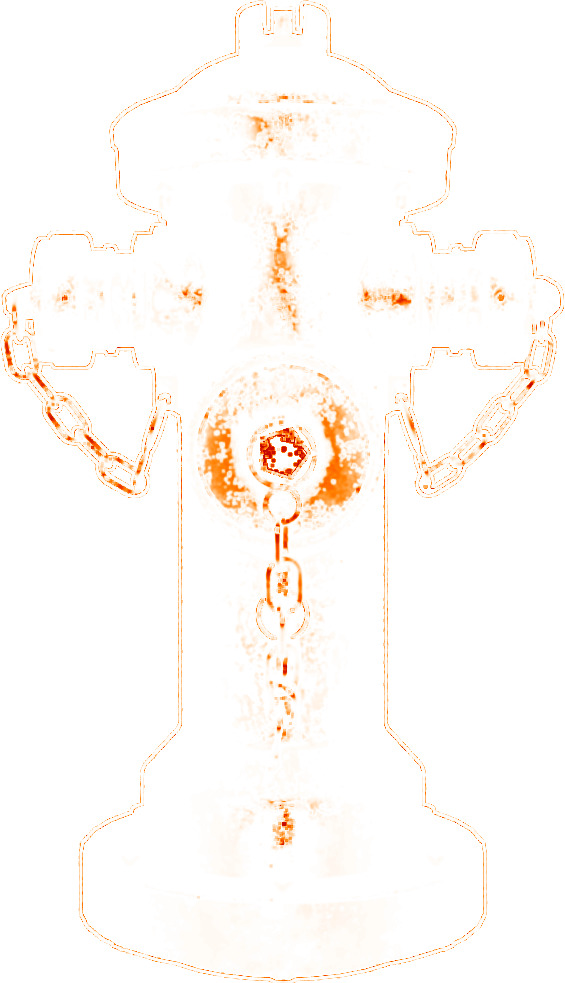}}
  \enskip\includegraphics[height=3.9cm]{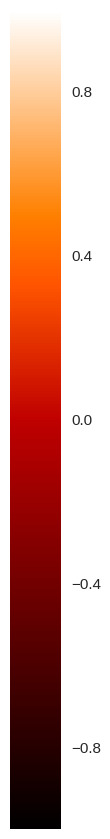}\qquad
  \includegraphics[height=3.9cm]{img/blender-ward2cycles-ggx/hydrant-blender-ward-fullrange-withdiffuse.jpg}
  \includegraphics[height=3.9cm]{img/blender-ward2cycles-ggx/hydrant-cycles-ggx-withdiffuse.jpg}
  \caption{\label{fig:chain-remapping-blender-ward-to-cycles-ggx}%
    Chained remapping from Ward (Blender) to GGX (Cycles) via two intermediate BRDF models.
    \emph{Left to right:} Blender internal-Ward, Mitsuba-Ward, Mitsuba-GGX, Blender
    Cycles-GGX, SSIM error. \emph{With diffuse term:} Blender internal-Ward, Blender Cycles-GGX.}
\end{figure*}

In general, we find the remapping scheme stable enough to enable chaining of
transformations. Figure~\ref{fig:chain-remapping-blender-ward-to-cycles-ggx} shows an
example transformation chain: $\text{Blender internal-Ward} \rightarrow
\text{Mitsuba-Ward} \rightarrow \text{Mitsuba-GGX} \rightarrow \text{Blender Cycles-GGX}$,
where the first transformation is the one depicted in
Figure~\ref{fig:blender-ward_vs_mitsuba-ward_plots}).

Apart from the inherently increased sheen due to GGX, the result remains remarkably close
to the input. Note that in spite of working with three different renderers, the most
noticeable differences occur when remapping from Ward to GGX within the same renderer
(Mitsuba). This hints that the BRDF model shape is the main factor determining the
remappability of a material, despite other additional differences that may be in play
between renderers (e.g., source light behaviour, post-processing, etc.). In principle, the
ability to chain transformations in this way allows to convert between a wide range of
BRDF models and renderers without having to determine individual transformation for all
possible pairs.

\subsection{Comparison with SVR}
\label{sec:results:svr}

\ASadd{As mentioned in Section~\ref{sec:parametric-scheme}, the mapping between BRDF parameters
can be learned using a regression method of general scope, without the need to make assumptions
on the functional shape of the mapping. For the purpose of comparison we tested
Support Vector Regression with radial basis functions, which requires an extra step
of optimisation of the hyper-parameters through gradient-descent. Once trained, the
SVR was able to correctly model and interpolate parameter values inside the region
sampled by the training dataset, with equivalent results to our parametric approach in
terms of SVBRDF remapping. The results of SVR remapping can be seen in
Figure~\ref{fig:blender-ward_vs_mitsuba-ward_renderings}, where all the material parameters
in the texture maps lie inside the subregion sampled by our dataset (i.e. the datapoints seen in
Figure~\ref{fig:blender-ward_vs_mitsuba-ward_plots}). However, when applying the SVR
remapping to a general SVBRDF material which does not meet this requirement, we find that
the SVR fails to extrapolate the behaviour of the transformation outside of the sample
space, thus generating fringe changes in the resulting asset.
In Figure~\ref{fig:parametric_vs_svr} we show the remapping of a spatially-varying material
by means of both parametric (bottom-center) and SVR (bottom-right) regressions. The
renderings use the environment map illumination from Figure~\ref{fig:envmap}.
\begin{figure}[h!]
	\centering
	\includegraphics[width=0.9\linewidth]{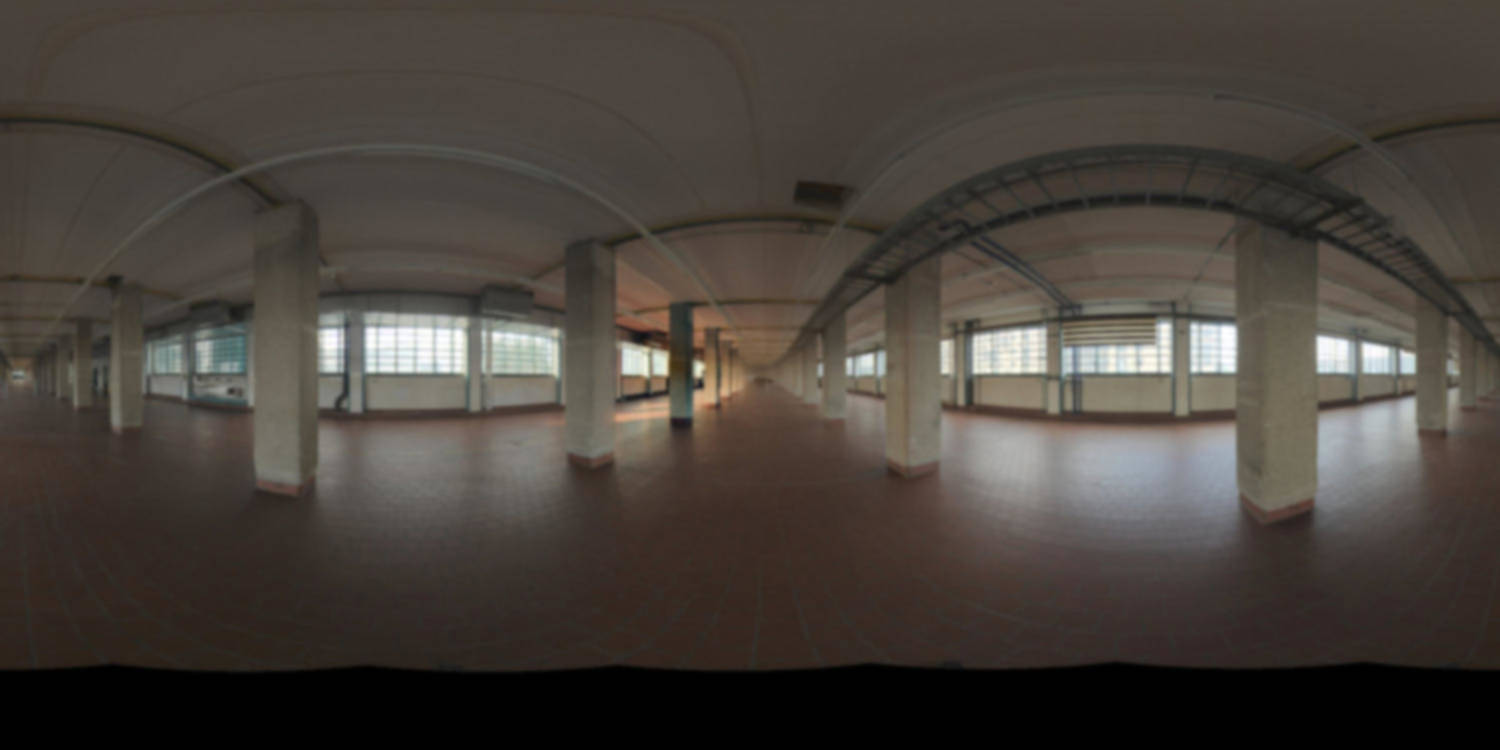}
	\caption{\label{fig:envmap}
	Tabac plant environment map.}
\end{figure}

Due to the extrapolation issues with SVR, the resulting remapping has suffered a very noticeable
change in chromaticity. In contrast, the functional shape of our parametric function
assumes a linear extrapolation of the specular values. This means that in the transformation
of parameters, the three channels of each texel are multiplied by the same factor,
thus leading to a conservation of the chromaticity.}

\begin{figure}[htb]
	\centering
	\includegraphics[height=4.5cm]{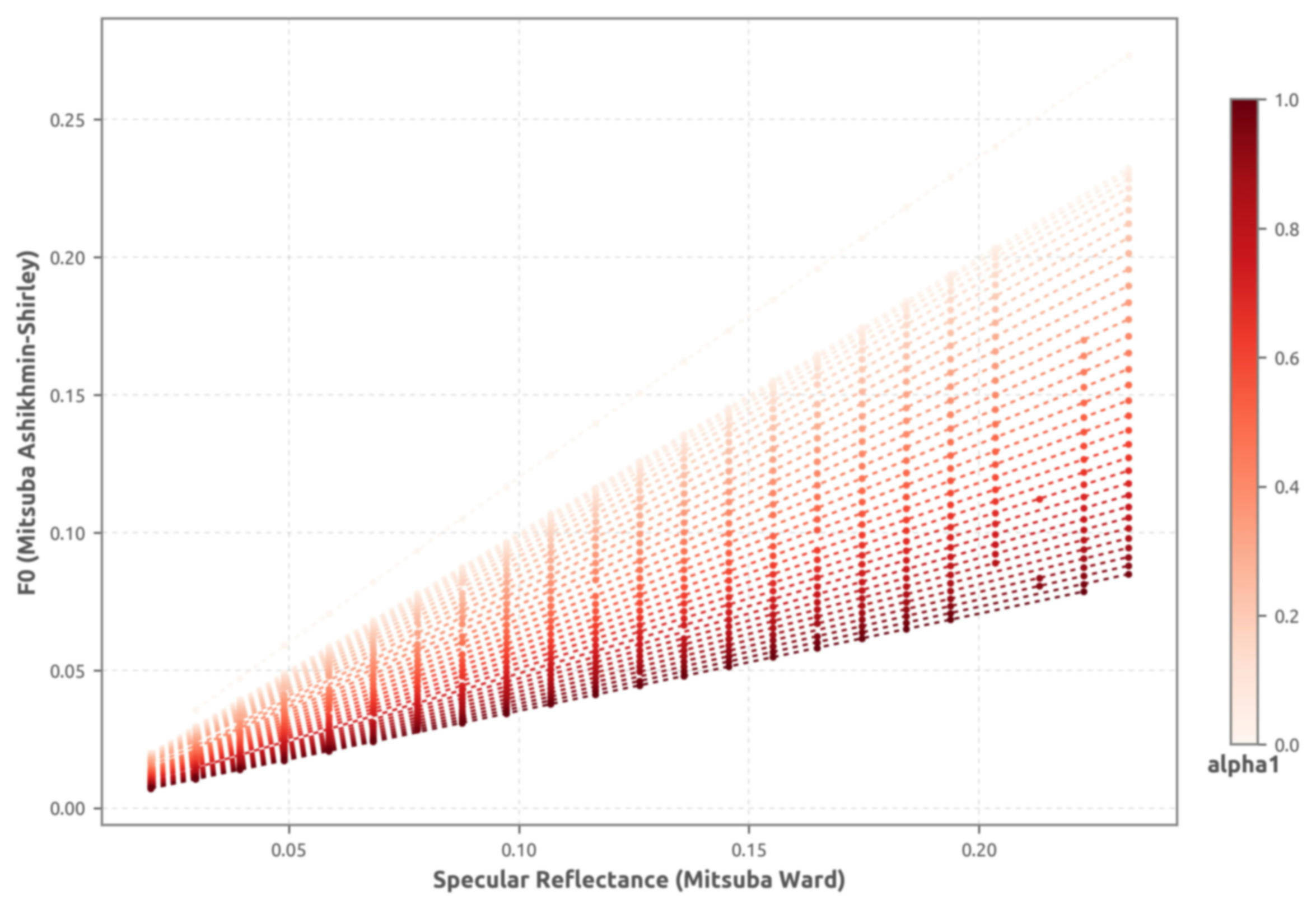}\\
	\includegraphics[height=4.5cm]{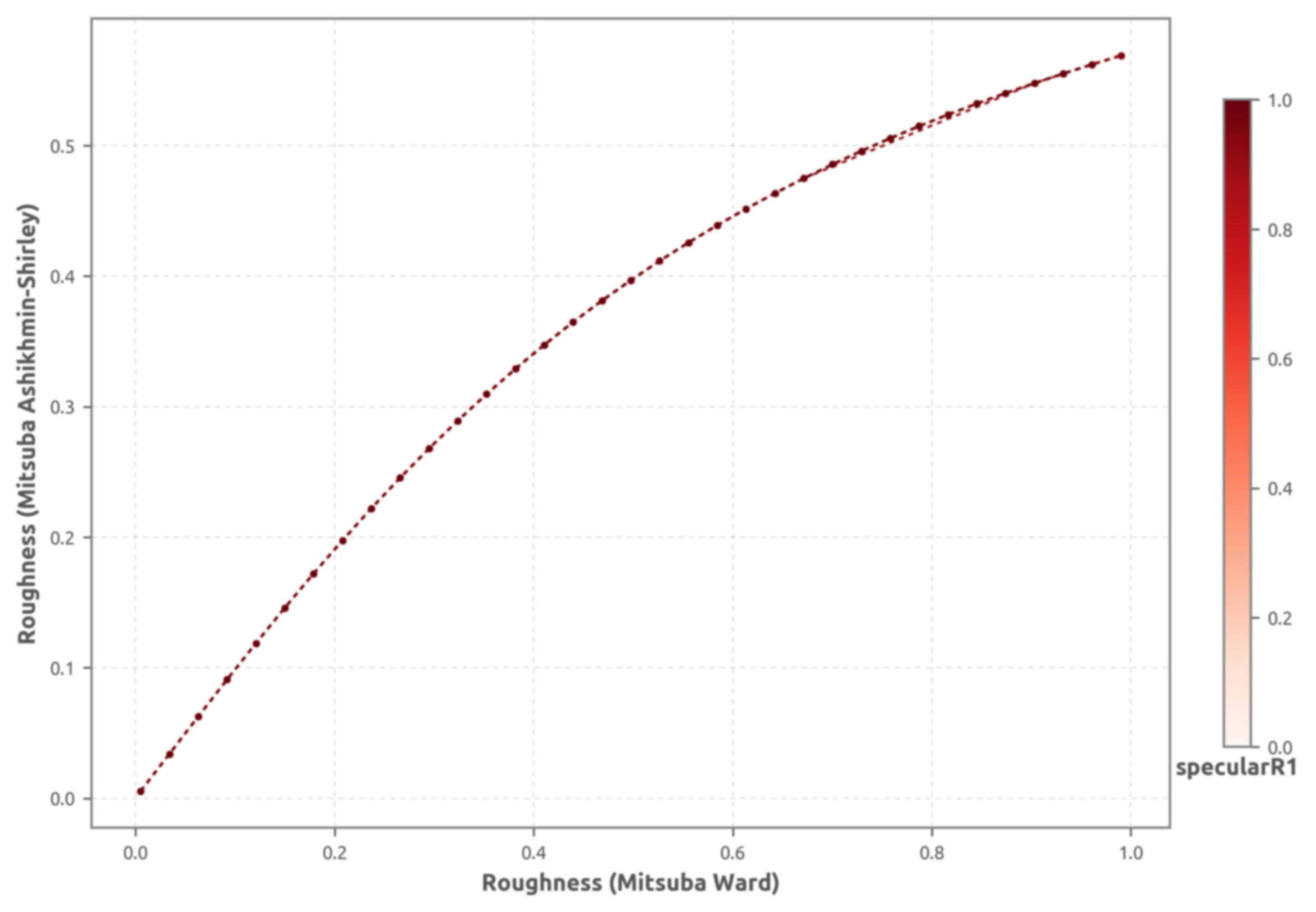}\\
	\vspace{1em}

	\includegraphics[height=3.5cm]{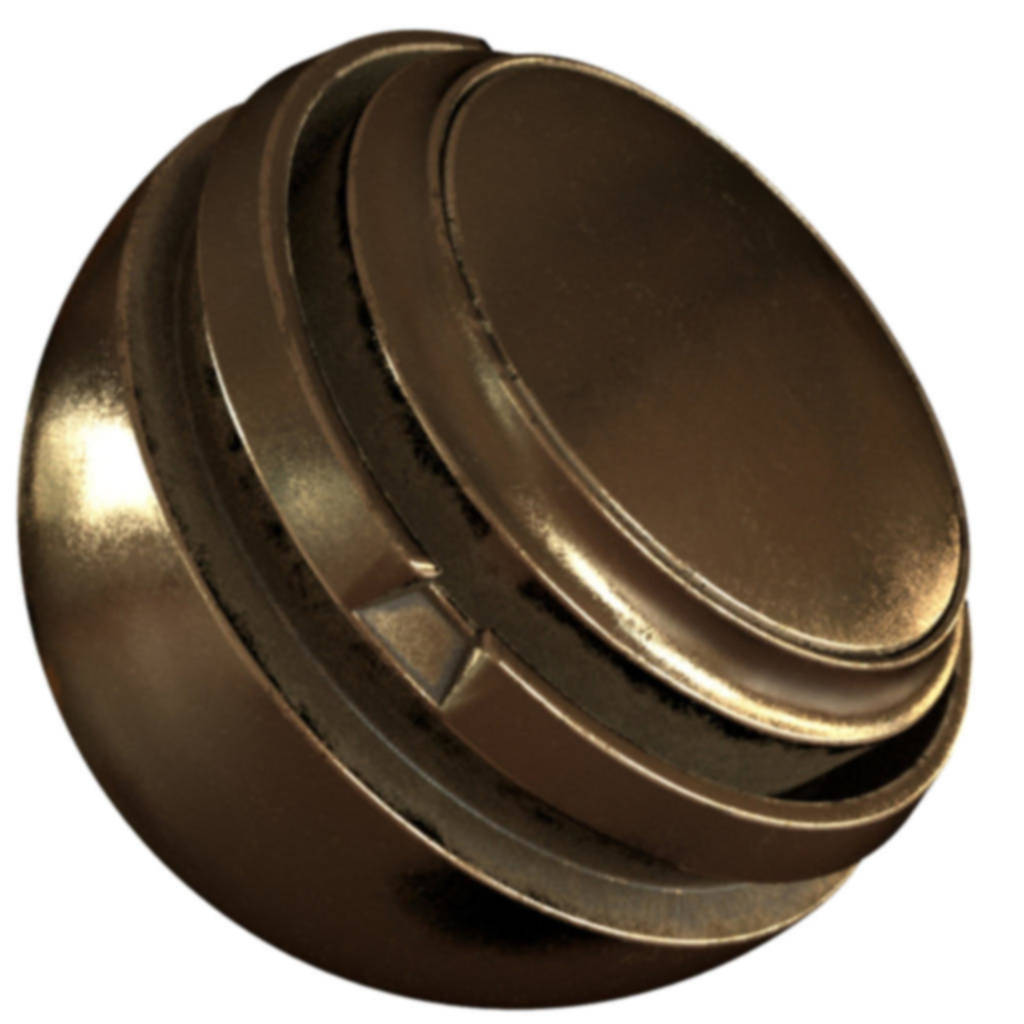}
	\includegraphics[height=3.5cm]{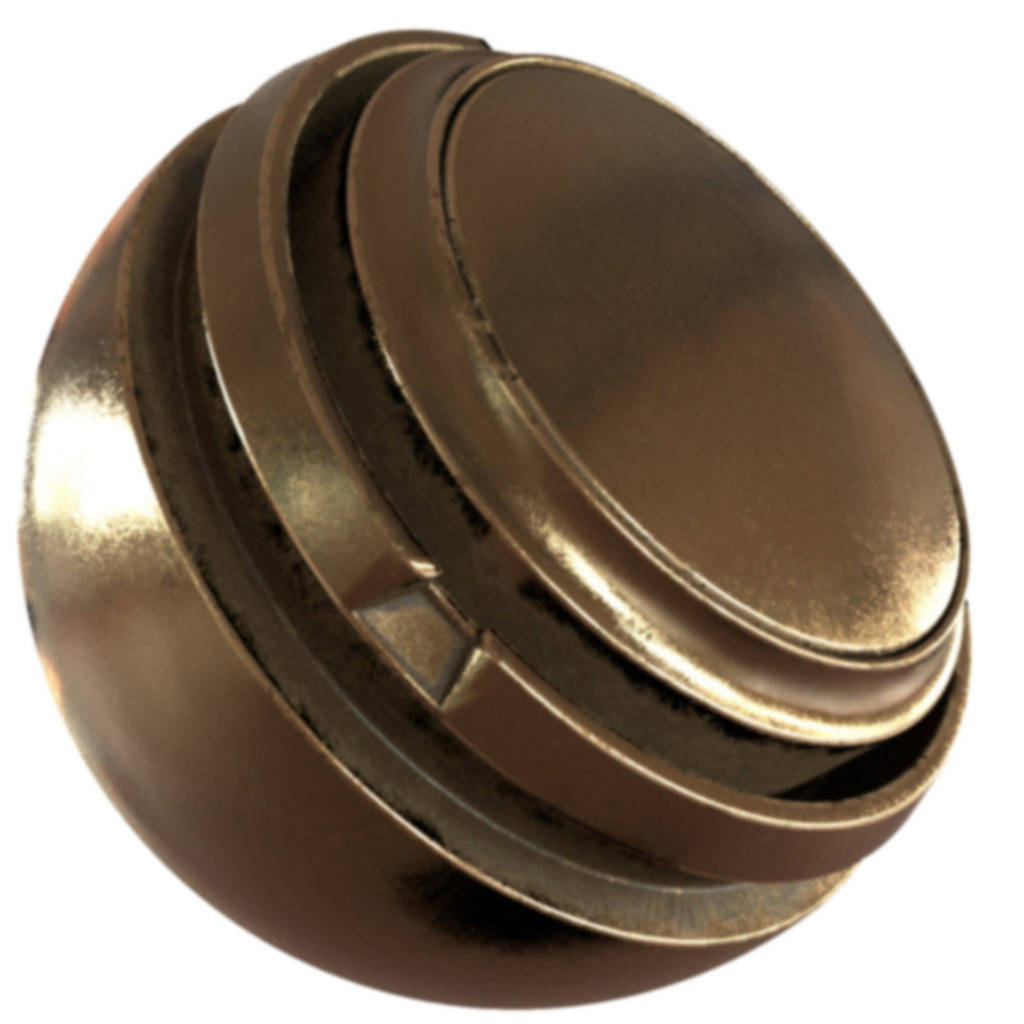}\\
	\includegraphics[height=3.5cm]{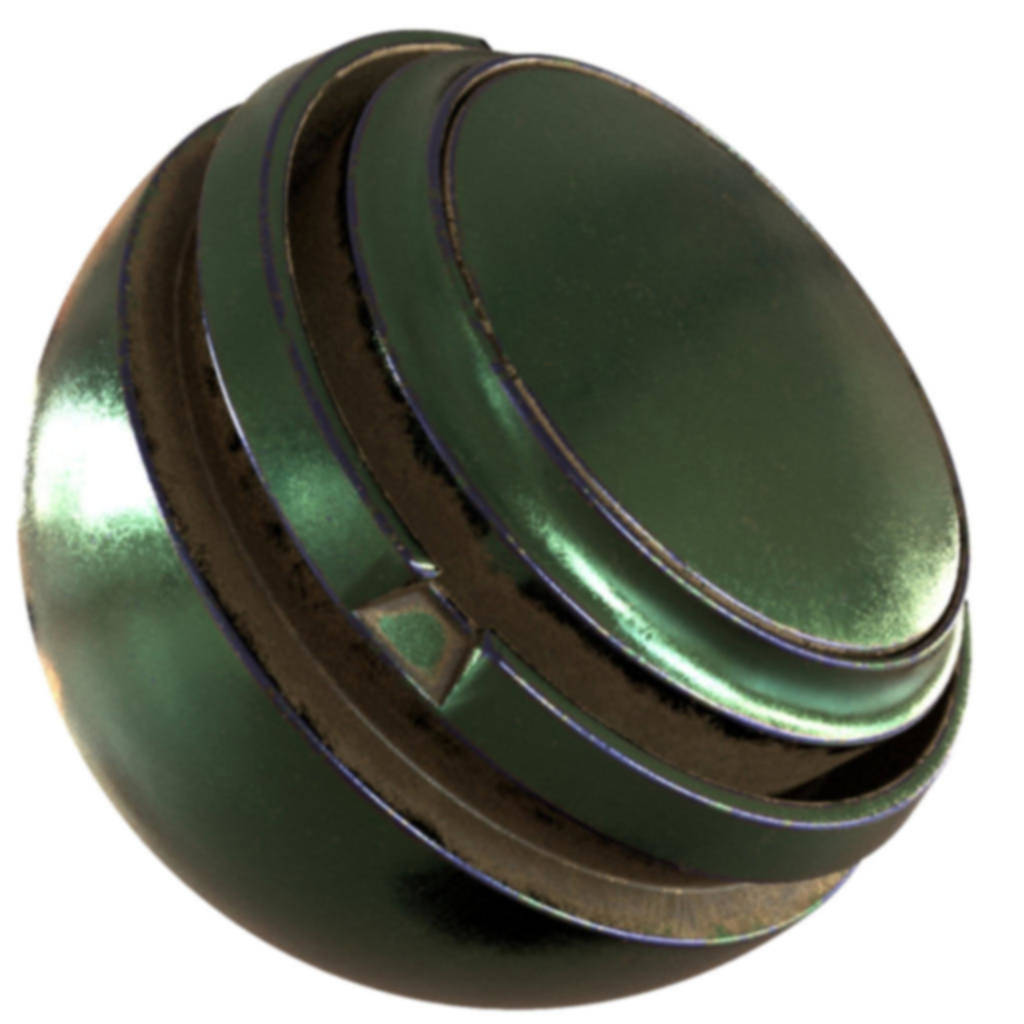}
	\caption{\label{fig:parametric_vs_svr}%
		\emph{Plots:} Remapping from Mitsuba Ward to Mitsuba Ashikhmin-Shirley. Remapping of
		specular reflectance with color indicating roughness, and remapping of roughness.
		\emph{Renderings:} Renderings of the remapping of a spatially-varying material (\textit{scratched gold}) using two different methods for regression. 
		Left: original Mitsuba-Ward. Right: remapped Mitsuba Ashikhmin-Shirley with our parametric scheme. Bottom: remapped Mitsuba Ashikhmin-Shirley with Support Vector Regression.
		Illumination is provided by the environment map from Figure~\ref{fig:envmap}.
	}
\end{figure}

\subsection{Illuminant position}
\label{sec:results:nonheadlight}

\begin{figure}[h!]
	\centering
	\includegraphics[width=0.9\linewidth]{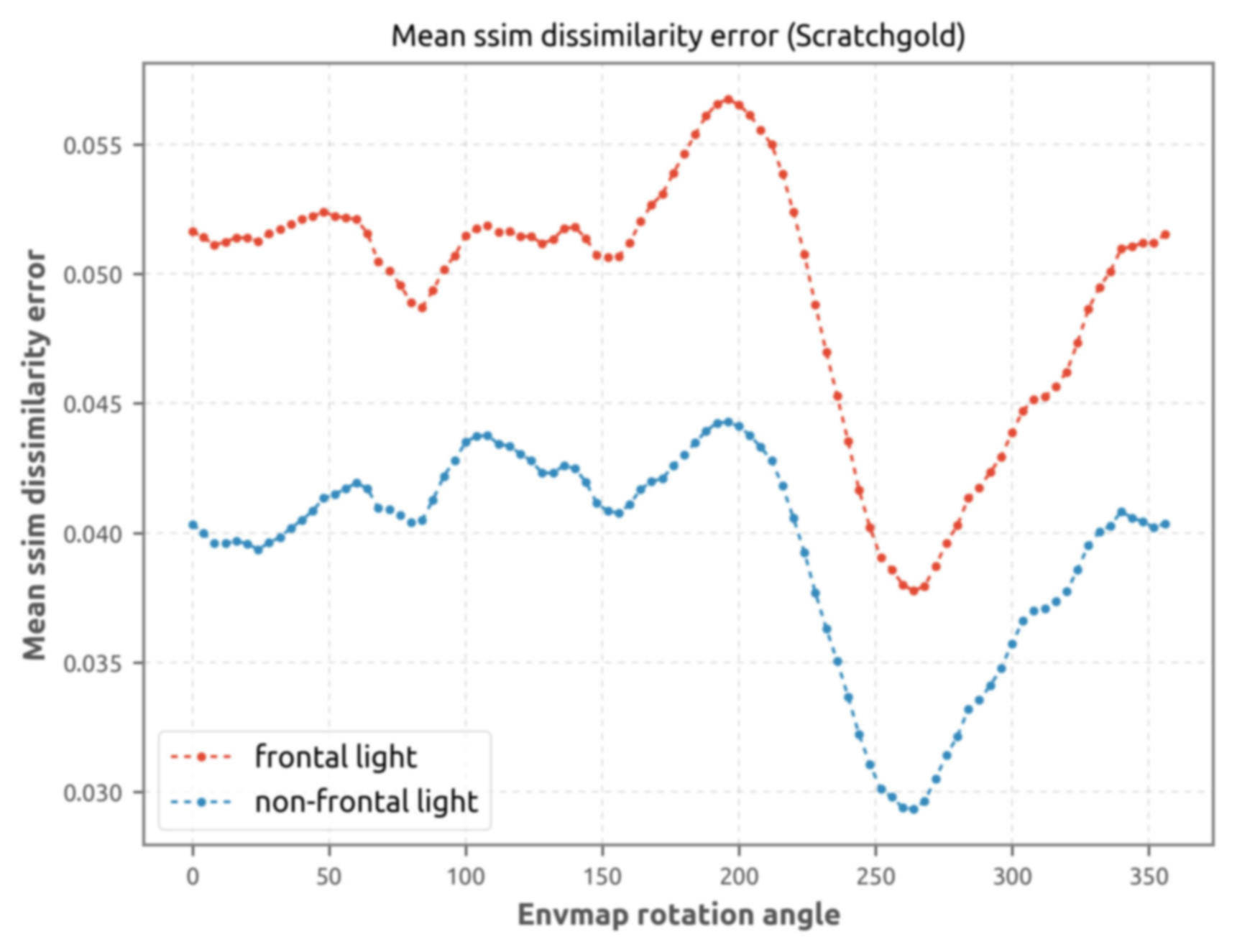}
	\includegraphics[width=0.9\linewidth]{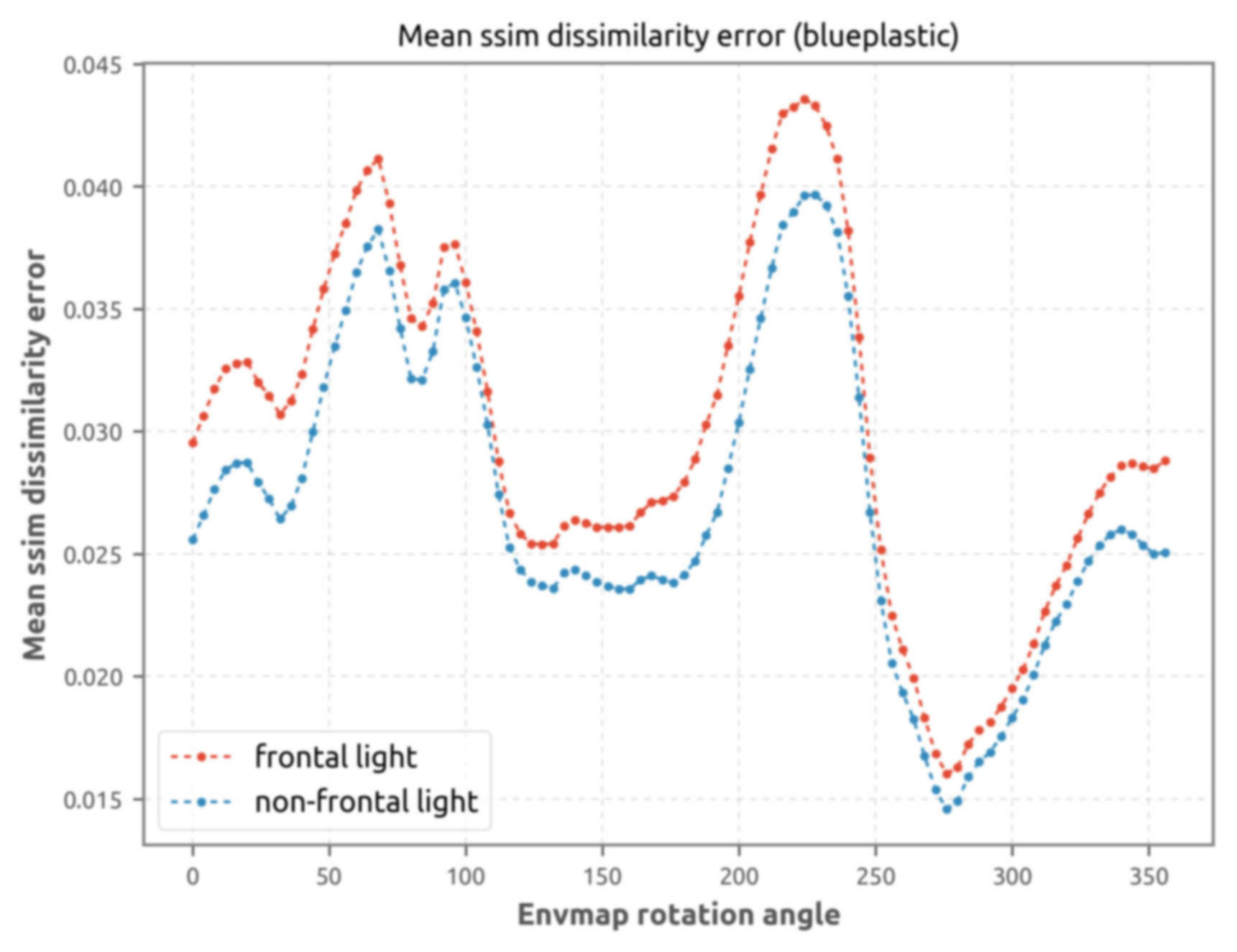}
	\includegraphics[width=0.9\linewidth]{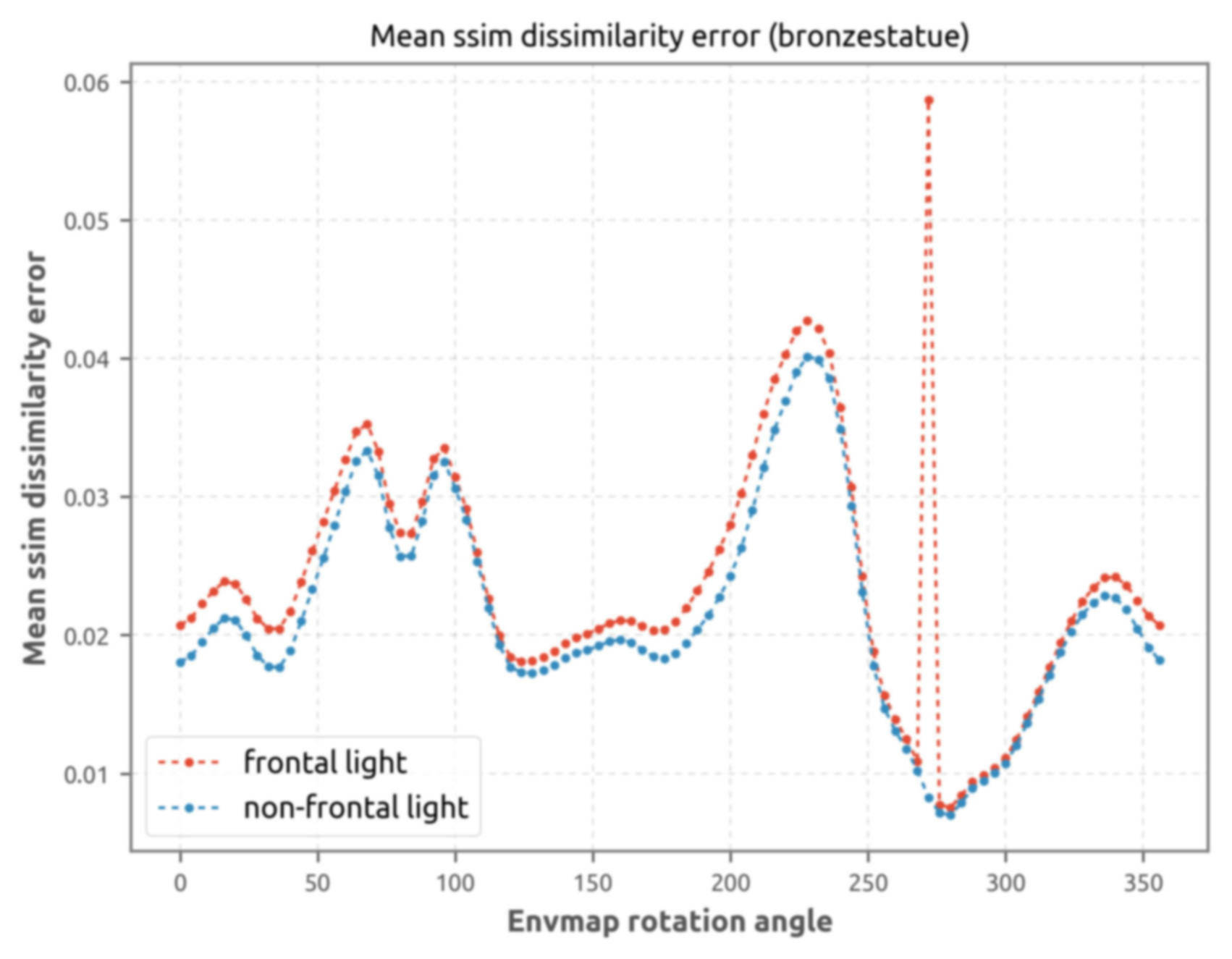}
	\caption{\label{fig:ssim-plots}
	Mean dissimilarity error ($1 - \text{SSIM}$) for three SVBRDF materials as a function of the
	rotation angle of the illuminating environment map from Figure~\ref{fig:envmap}.
	Higher values indicate higher error.
	\emph{Top:} scratched gold. \emph{Center:} blue plastic. \emph{Bottom:} Bronze statue.
	Remapping scene with headlight (red) and non-headlight illumination (blue).
	}
\end{figure}

\begin{figure*}[h!]
  \centering
  \includegraphics[height=0.12\textheight]{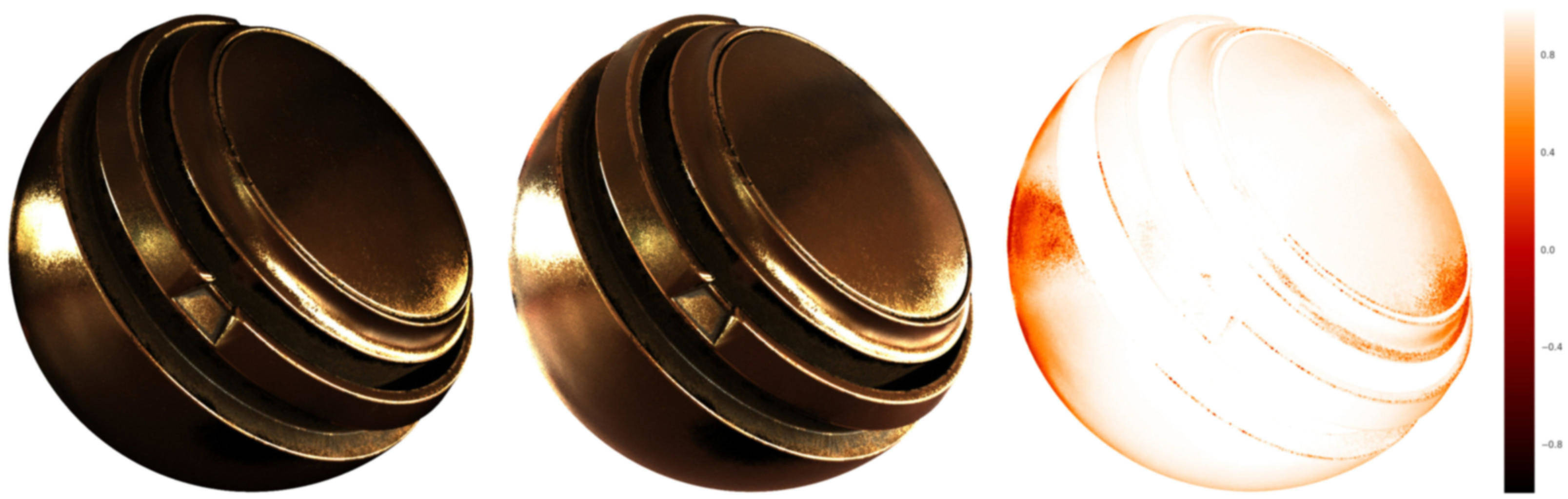}~
  \includegraphics[height=0.12\textheight]{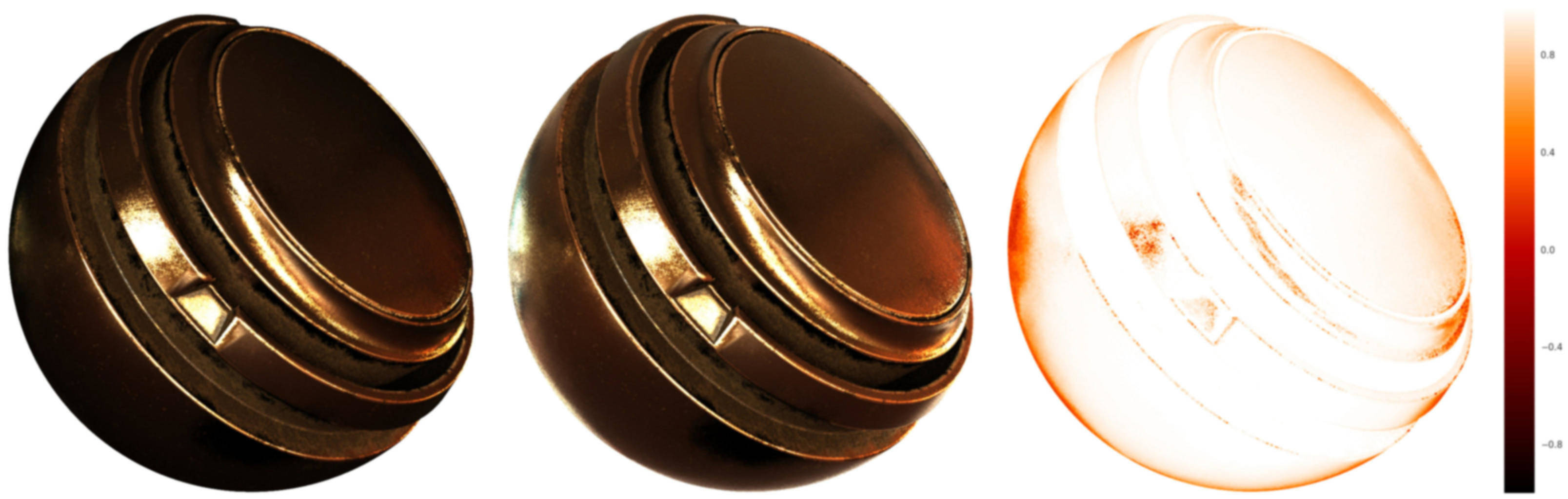}\\
  \includegraphics[height=0.12\textheight]{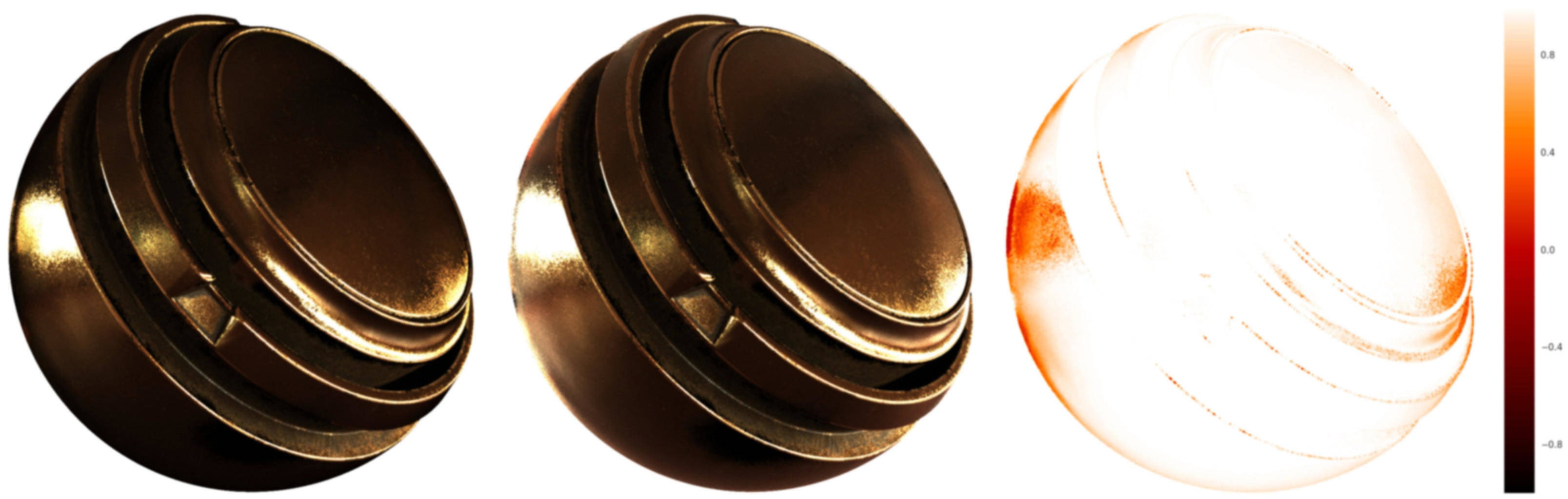}~
  \includegraphics[height=0.12\textheight]{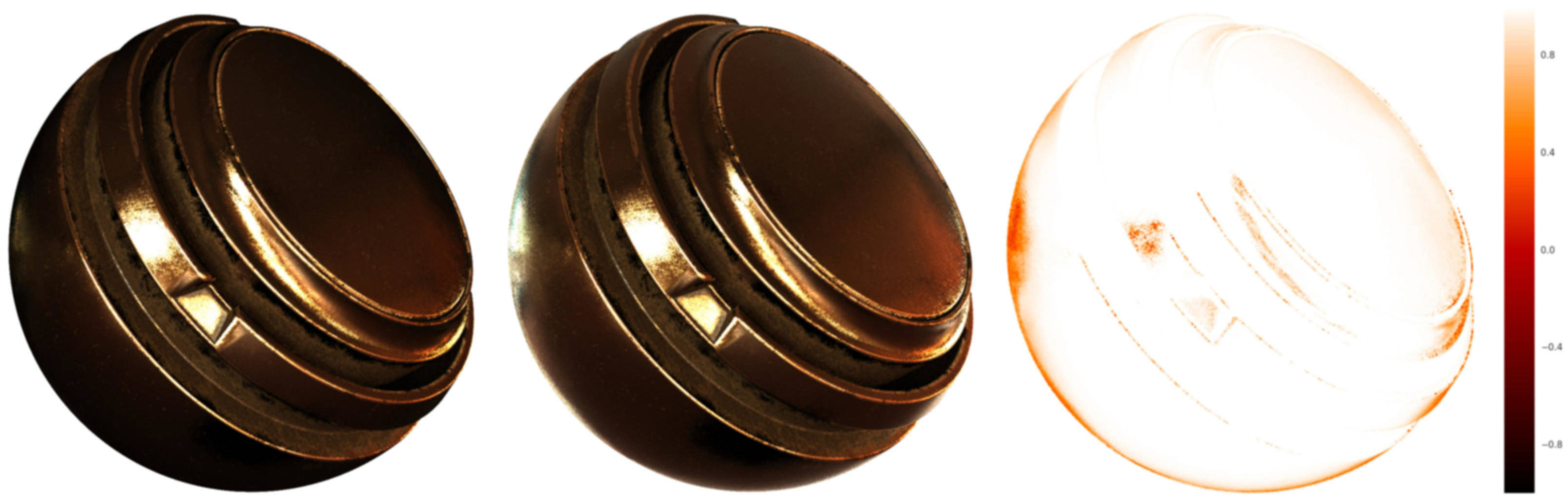}
  \caption{\label{fig:scratchgold_front_diag}
  Remappings of scratched gold with frontal (top) and non-frontal light (bottom).
  Illumination provided by environment map from Figure~\ref{fig:envmap} at rotation angles
  $0^\circ$ (left side) and $272^\circ$ (right side).
  Within each side: \emph{Left:} original material in Mitsuba Ward. \emph{Center:} remapped in Mitsuba Ashikhmin-Shirley. \emph{Right:} SSIM difference. 
  }
\end{figure*}

\begin{figure*}[h!]
  \centering
  \includegraphics[height=0.12\textheight]{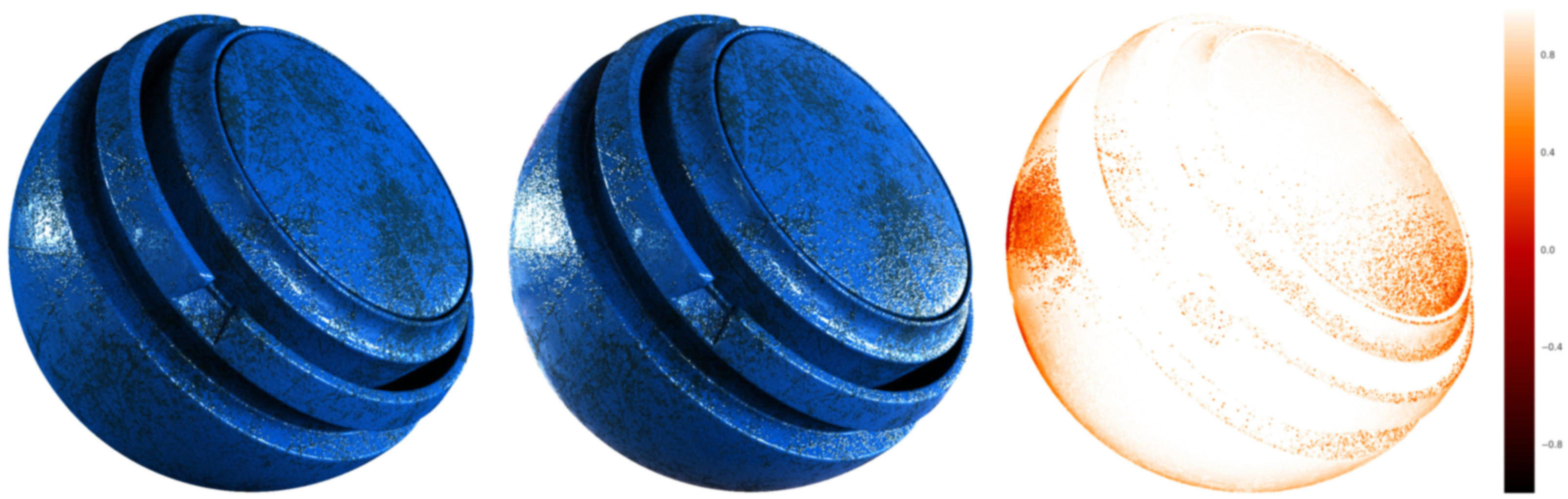}~
  \includegraphics[height=0.12\textheight]{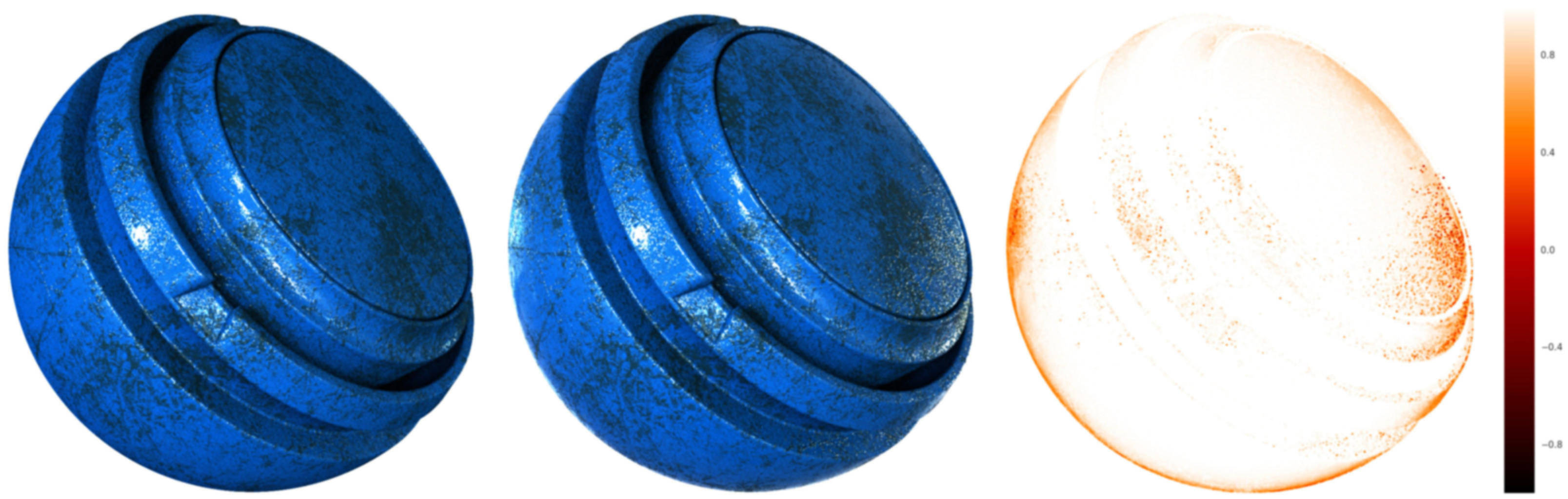}\\
  \includegraphics[height=0.12\textheight]{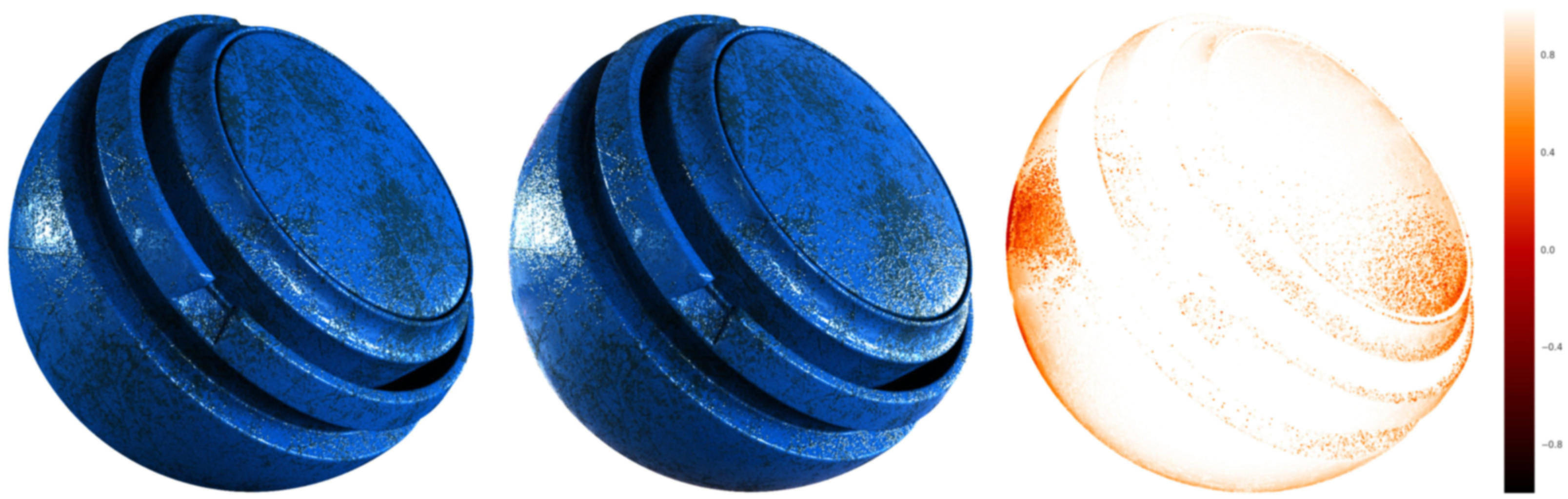}~
  \includegraphics[height=0.12\textheight]{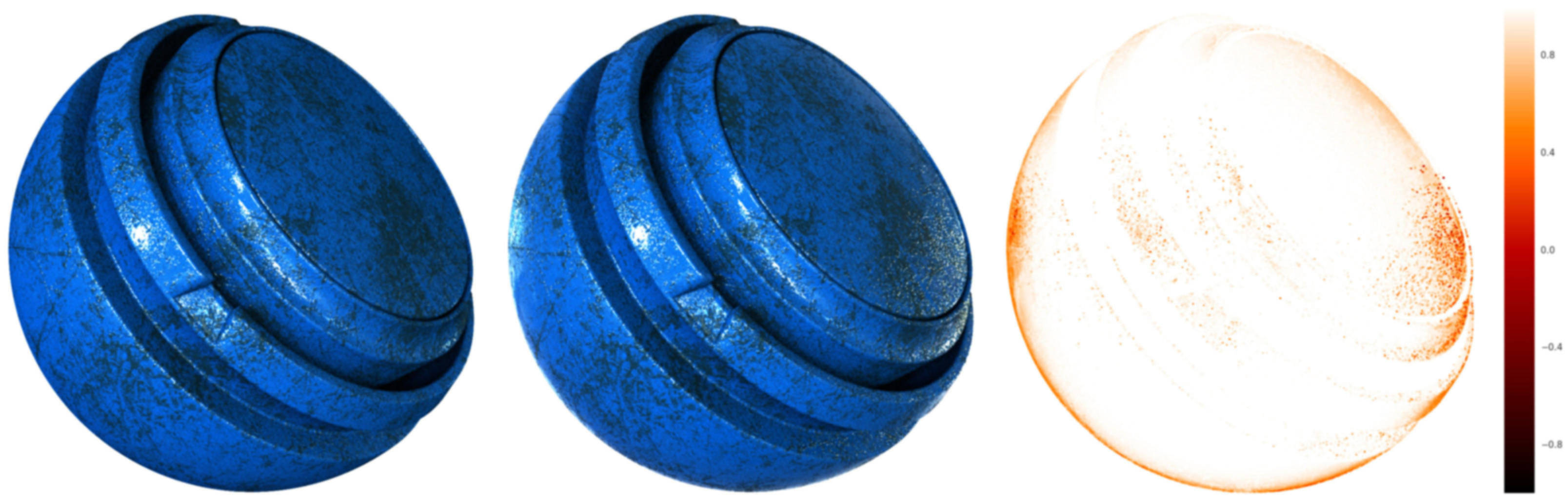}
  \caption{\label{fig:blueplastic_front_diag}
  Remappings of blue plastic with frontal (top) and non-frontal light (bottom).
  Illumination provided by environment map from Figure~\ref{fig:envmap} at rotation angles
  $0^\circ$ (left side) and $272^\circ$ (right side).
  Within each side: \emph{Left:} original material in Mitsuba Ward. \emph{Center:} remapped in Mitsuba Ashikhmin-Shirley. \emph{Right:} SSIM difference. 
  }
\end{figure*}

\begin{figure*}[h!]
	\centering
	\includegraphics[height=0.12\textheight]{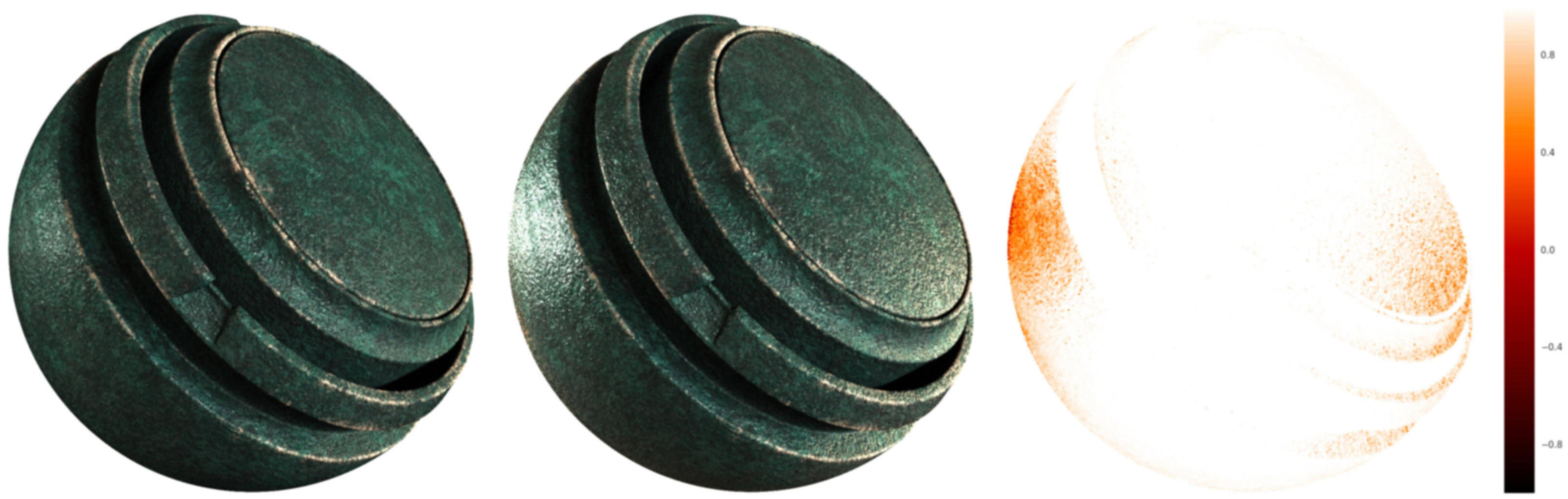}~
	\includegraphics[height=0.12\textheight]{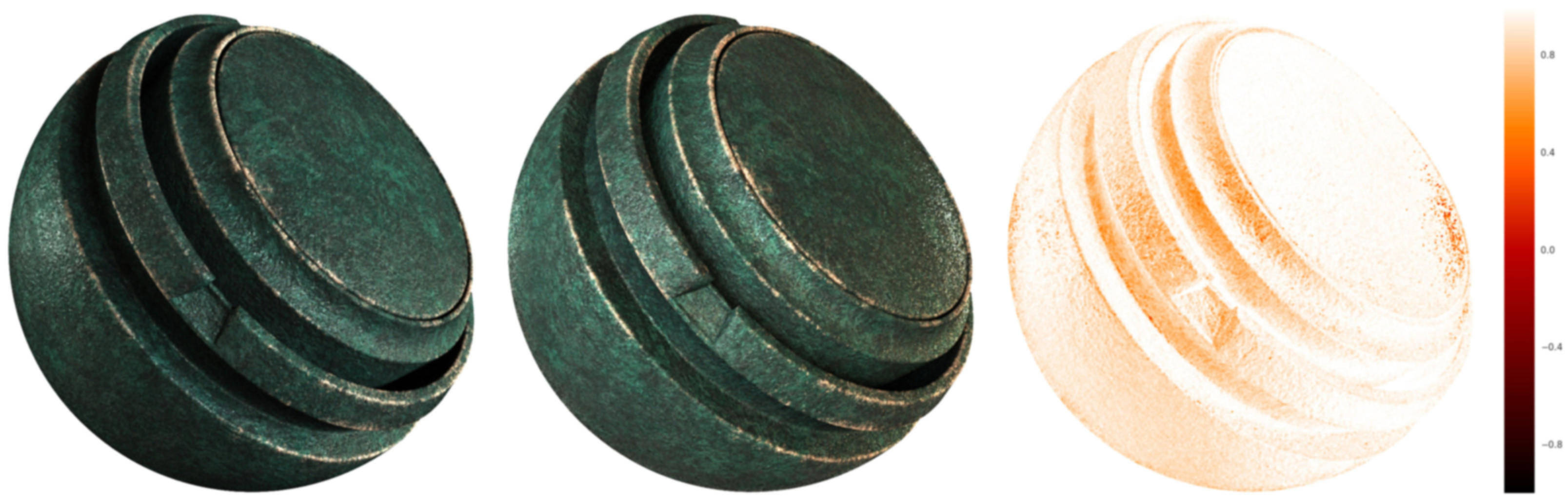}\\
	\includegraphics[height=0.12\textheight]{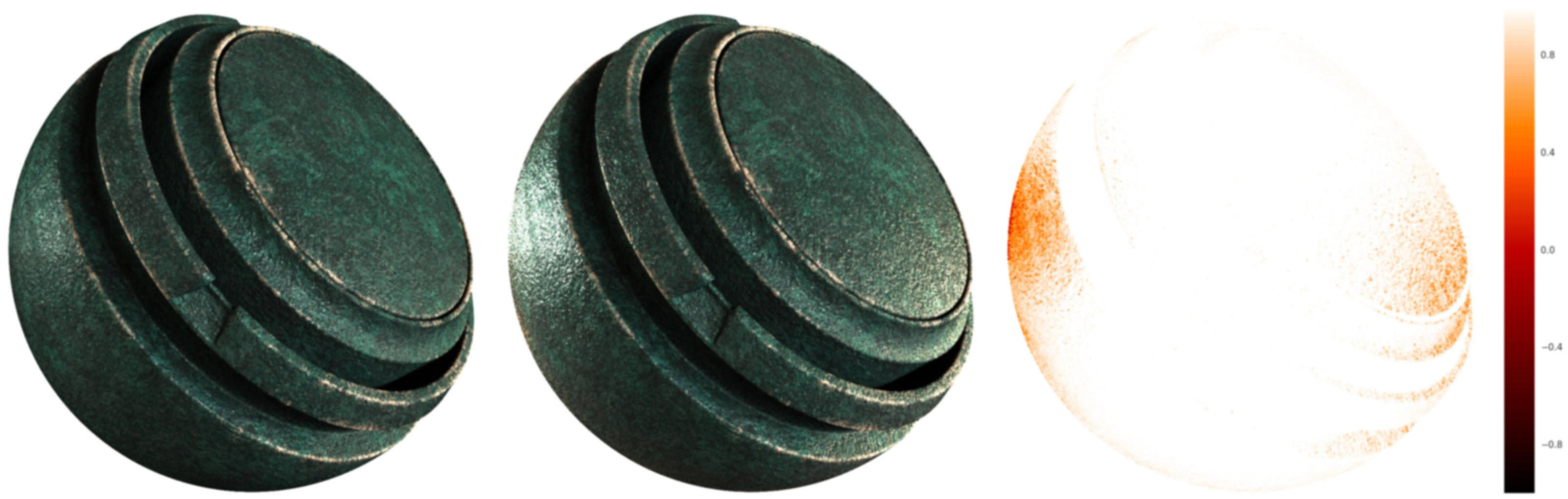}~
	\includegraphics[height=0.12\textheight]{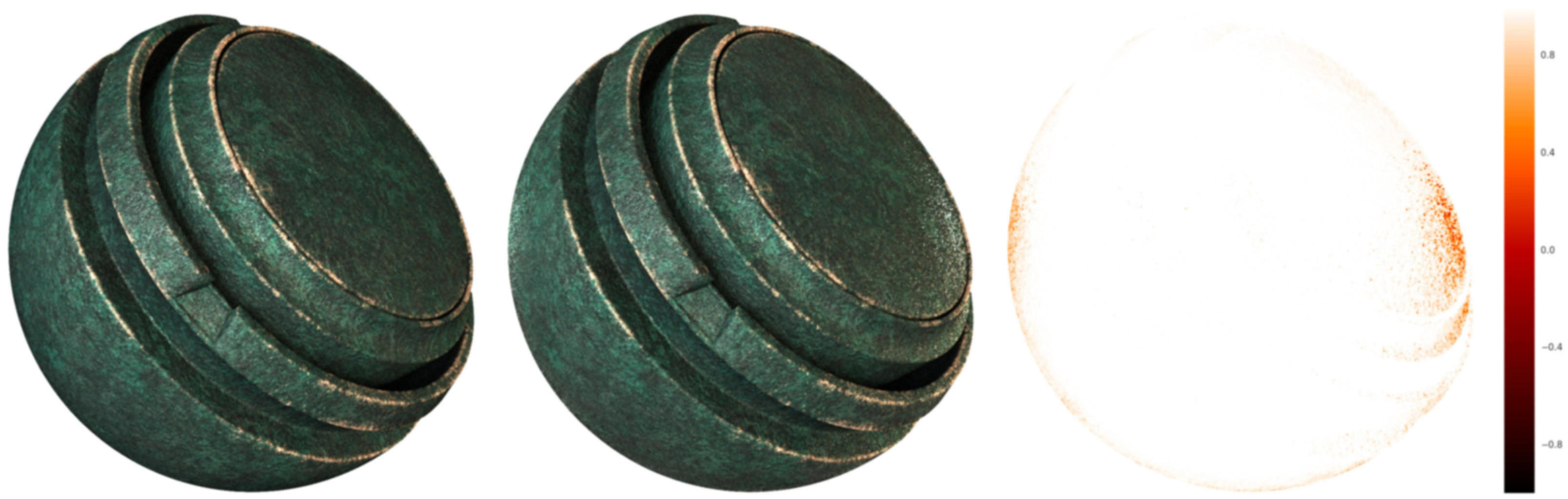}
	\caption{\label{fig:bronzestatue_front_diag}
  Remappings of bronze statue with frontal (top) and non-frontal light (bottom).
  Illumination provided by environment map from Figure~\ref{fig:envmap} at rotation angles
  $0^\circ$ (left side) and $272^\circ$ (right side).
  Within each side: \emph{Left:} original material in Mitsuba Ward. \emph{Center:} remapped in Mitsuba Ashikhmin-Shirley. \emph{Right:} SSIM difference. 
	}
\end{figure*}

\ASadd{
As previously discussed in Section~\ref{sec:results:uniform}, our scheme for uniform BRDF
remapping utilises a rendered scene with point light illumination to provide a partial
sampling of the material's reflectance. The main requirement that we found for the light
position during our experiments is that a great proportion of the pixels in the resulting
renderings should be illuminated, so that the optimisation is able to converge and we
obtain a stable transformation. This needs to happen for all considered combinations of
parameters, including materials with low roughness where the highlight does not spread
far from the direction of specular reflection. An obvious choice for this purpose is a
lightsource concentric with the camera, which maximises the size of the specular
highlight in the rendering, thus improving the characterisation of the lobe. However,
the symmetry of the scene configuration poses multiple potential issues: (1) an over-representation
of the retroreflective lobe in the sampling; (2) a repeated sampling of directions
which in isotropic BRDFs are equivalent and do not provide new information; (3) for many
BRDF models, insufficient coverage of the parameter space to constrain all parameters,
for instance the Fresnel term.

In Figures~\ref{fig:scratchgold_front_diag}-\ref{fig:bronzestatue_front_diag} we display
the remappings of spatially-varying materials with two different light settings: frontal
light (top) and non-frontal light with $\theta_l = 45^\circ$ (bottom). Visual differences
between these two settings
for remapping are hard to spot, but in some cases a slight decrease of the global
dissimilarity is aparent when using non-frontal light. This is confirmed by the plots
in Figure~\ref{fig:ssim-plots} where we can
observe the corresponding mean dissimilarity errors of these materials as we rotate
the environment map illumination (note that we refer to a rotation of the illumination
used for the renderings, \textit{not} the one used in the remapping). Videos of these
and other remappings with rotating environment illumination can be found in the supplemental
material. \AS{Plots of absolute and relative error show exactly the same kind of behaviour
as these SSIM plots.}
}

\section{Conclusions}

We presented a method for automatic translation of material appearance
between different BRDF models and across different renderers, which uses an image-based metric for
appearance comparison, and that delegates the interaction with the
model to the renderer. We analysed the performance of the method, both
with respect to robustness and visual differences of the fits for
multiple combinations of BRDF models. While it is effective for
individual BRDFs, the computational cost does not scale well for
spatially-varying BRDFs. Therefore, \ASadd{we also presented an interpolation scheme
based on a non-linear parametric regression of the transformation between BRDF model.
We used this to generate a
reduced polynomial representation of the transformation which
evaluates instantly and without further interaction with the
renderer, allowing the remapping of SVBRDF texture maps.} Moreover, the resulting transformations lend themselves to
chaining, enabling effortless transitions between BRDF models.
\ASadd{We compared our regression scheme with Support Vector Regression and showed that it provides
a better extrapolation of model parameters outside of the area defined by the training data.
Finally we analysed the effect of the lighting used during uniform remapping in the quality
of renderings of spatially-varying materials, and confirmed that headlight illumination 
leads to a slight increase in visual differences.}

\bibliographystyle{eg-alpha} 
\bibliography{eg}

\newcommand{\etalchar}[1]{$^{#1}$}
\begin{thebibliography}{\uppercase{WMLT07}}

\bibitem[All17]{allegorithmic2017}
Allegorithmic: Substance designer and substance painter.
\newblock \url{https://www.allegorithmic.com}, 2017.
\newblock Last access Oct.~2017.

\bibitem[AS00]{ashikhmin2000anisotropic}
\textsc{Ashikhmin M., Shirley P.}:
\newblock An anisotropic phong brdf model.
\newblock \emph{J. Graph. Tools 5}, 2 (Feb. 2000), 25--32.

\bibitem[BLPW14]{brady2014genbrdf}
\textsc{Brady A., Lawrence J., Peers P., Weimer W.}:
\newblock {genBRDF}: Discovering new analytic {BRDFs} with genetic programming.
\newblock \emph{ACM Trans. Graph. 33}, 4 (July 2014), 114:1--114:11.

\bibitem[Bur12]{burley2012physically}
\textsc{Burley B.}:
\newblock Physically-based shading at disney.
\newblock In \emph{ACM SIGGRAPH 2012 Courses} (2012), SIGGRAPH `12.

\bibitem[Cor17]{corona2017}
Corona renderer.
\newblock \url{https://corona-renderer.com}, 2017.
\newblock Last access Oct.~2017.

\bibitem[Dam17]{cyrille.comm}
\textsc{Damez C.}:
\newblock {Personal communication (CTO, Allegorithmic)}, Oct.~2017.

\bibitem[DHT{\etalchar{*}}00]{debevec2000acquiring}
\textsc{Debevec P., Hawkins T., Tchou C., Duiker H.-P., Sarokin W., Sagar M.}:
\newblock Acquiring the reflectance field of a human face.
\newblock In \emph{Proceedings of the 27th Annual Conference on Computer
  Graphics and Interactive Techniques} (New York, NY, USA, 2000), SIGGRAPH '00,
  ACM Press/Addison-Wesley Publishing Co., pp.~145--156.

\bibitem[GGG{\etalchar{*}}16]{guarnera2016brdf}
\textsc{Guarnera D., Guarnera G., Ghosh A., Denk C., Glencross M.}:
\newblock {BRDF} representation and acquisition.
\newblock \emph{Comput. Graph. Forum 35}, 2 (May 2016), 625--650.

\bibitem[HFM16]{havran2016}
\textsc{Havran V., Filip J., Myszkowski K.}:
\newblock {Perceptually Motivated {BRDF} Comparison using Single Image}.
\newblock \emph{Computer Graphics Forum} (2016).

\bibitem[Jak10]{mitsuba}
\textsc{Jakob W.}:
\newblock Mitsuba renderer, 2010.
\newblock http://www.mitsuba-renderer.org.

\bibitem[Kri17]{jaroslav.comm}
\textsc{Krivanek J.}:
\newblock {Personal communication (Head of Research, Corona Renderer)},
  Jul.~2017.

\bibitem[LFTG97]{lafortune1997}
\textsc{Lafortune E. P.~F., Foo S.-C., Torrance K.~E., Greenberg D.~P.}:
\newblock Non-linear approximation of reflectance functions.
\newblock In \emph{Proceedings of the 24th Annual Conference on Computer
  Graphics and Interactive Techniques} (New York, NY, USA, 1997), SIGGRAPH '97,
  ACM Press/Addison-Wesley Publishing Co., pp.~117--126.

\bibitem[Mar98]{marschner1998inverse}
\textsc{Marschner S.~R.}:
\newblock \emph{Inverse Rendering for Computer Graphics}.
\newblock PhD thesis, Cornell University, Ithaca, NY, 1998.

\bibitem[May17]{maya2017}
Autodesk maya.
\newblock \url{https://www.autodesk.com/products/maya}, 2017.
\newblock Last access Oct.~2017.

\bibitem[NDM05]{ngan2005}
\textsc{Ngan A., Durand F., Matusik W.}:
\newblock Experimental analysis of {BRDF} models.
\newblock In \emph{Proceedings of the Sixteenth Eurographics Conference on
  Rendering Techniques} (Aire-la-Ville, Switzerland, Switzerland, 2005), EGSR
  '05, Eurographics Association, pp.~117--126.

\bibitem[NDM06]{ngan2006image}
\textsc{Ngan A., Durand F., Matusik W.}:
\newblock Image-driven navigation of analytical brdf models.
\newblock In \emph{Proceedings of the 17th Eurographics Conference on Rendering
  Techniques} (Aire-la-Ville, Switzerland, Switzerland, 2006), EGSR '06,
  Eurographics Association, pp.~399--407.

\bibitem[PJH16]{pharr2016pbr}
\textsc{Pharr M., Jakob W., Humphreys G.}:
\newblock \emph{Physically Based Rendering, Third Edition: From Theory To
  Implementation}, 3rd~ed.
\newblock Morgan Kaufmann Publishers Inc., San Francisco, CA, USA, 2016.

\bibitem[SDSG13]{schregle2013}
\textsc{Schregle R., Denk C., Slusallek P., Glencross M.}:
\newblock {Grand Challenges: Material Models in the Automotive Industry}.
\newblock In \emph{Eurographics Workshop on Material Appearance Modeling}
  (2013), Klein R., Rushmeier H., (Eds.), The Eurographics Association.

\bibitem[Uni17]{unity2017}
Unity.
\newblock \url{https://unity3d.com/}, 2017.
\newblock Last access Oct.~2017.

\bibitem[Unr17]{unreal2017}
Unreal engine.
\newblock \url{https://www.unrealengine.com}, 2017.
\newblock Last access Oct.~2017.

\bibitem[WLL{\etalchar{*}}08]{weyrich2008principles}
\textsc{Weyrich T., Lawrence J., Lensch H., Rusinkiewicz S., Zickler T.}:
\newblock Principles of appearance acquisition and representation.
\newblock \emph{Foundations and Trends in Computer Graphics and Vision 4}, 2
  (2008), 75--191.

\bibitem[WMLT07]{walter2007microfacet}
\textsc{Walter B., Marschner S.~R., Li H., Torrance K.~E.}:
\newblock Microfacet models for refraction through rough surfaces.
\newblock In \emph{Proceedings of the 18th Eurographics Conference on Rendering
  Techniques} (Aire-la-Ville, Switzerland, Switzerland, 2007), EGSR'07,
  Eurographics Association, pp.~195--206.

\bibitem[WMP{\etalchar{*}}06]{weyrich2006analysis}
\textsc{Weyrich T., Matusik W., Pfister H., Bickel B., Donner C., Tu C.,
  McAndless J., Lee J., Ngan A., Jensen H.~W., Gross M.}:
\newblock Analysis of human faces using a measurement-based skin reflectance
  model.
\newblock \emph{ACM Trans. Graph. (Proc. SIGGRAPH) 25}, 3 (July 2006),
  1013--1024.

\end{thebibliography}

\end{document}